\documentclass[12pt,preprint]{aastex}
\usepackage{apjfonts}
\usepackage{epsfig}
\usepackage{psfig}
\usepackage{natbib}

\newcommand{\perval}[2]{{#1\mbox{$^{#2}$}}}

\newcommand{\degree}{{\rm o}}

\newcommand{\ee}[1]{\mbox{$10^{#1}$}}

\shorttitle{{X-ray Counterparts of Millisecond Pulsars in Globular Clusters}}
\shortauthors{W.~Becker, H.H.~Huang,\& T.~Prinz}

\begin{document}
\title{{X-ray Counterparts of Millisecond Pulsars in Globular Clusters}}
\author{ W.~Becker\altaffilmark{1}, H.H.~Huang\altaffilmark{1}, T.~Prinz\altaffilmark{1} }

\altaffiltext{1}{Max-Planck Institut f\"ur Extraterrestrische Physik, 85741 Garching bei M\"unchen, Germany}

\begin{abstract}
 We have systematically studied the X-ray emission properties of globular cluster  
 millisecond pulsars in order to evaluate their spectral properties and luminosities 
 in a uniform way. Cross-correlating the radio timing positions of the cluster 
 pulsars with the high resolution Chandra images revealed 31 X-ray counterparts 
 identified in nine different globular cluster systems, including those in 47\,Tuc.
 Timing analysis has been performed for all sources corresponding to the temporal 
 resolution available in the archival Chandra data. Making use of unpublished data 
 on M28, M4 and NGC 6752 allowed us to obtain further constraints for the millisecond 
 pulsar counterparts located in these clusters. Counting rate and energy flux upper 
 limits were computed for those 36 pulsars for which no X-ray counterparts could be 
 detected. Comparing the X-ray and radio pulse profiles of PSR J1821-2452 in M28 
 and the 47 Tuc pulsars PSR J0024-7204D,O,R indicated some correspondence between 
 both wavebands. The X-ray efficiency of the globular cluster millisecond pulsars 
 was found to be in good agreement with the efficiency $L_X\sim 10^{-3}\dot{E}$ 
 observed in Galactic field rotation-powered pulsars. Millisecond pulsars in the 
 galactic plane and in globular clusters appear to show no distinct differences 
 in their X-ray emission properties.
\end{abstract}


\section{INTRODUCTION \label{intro}}

\begin{figure}
\centerline{\psfig{figure=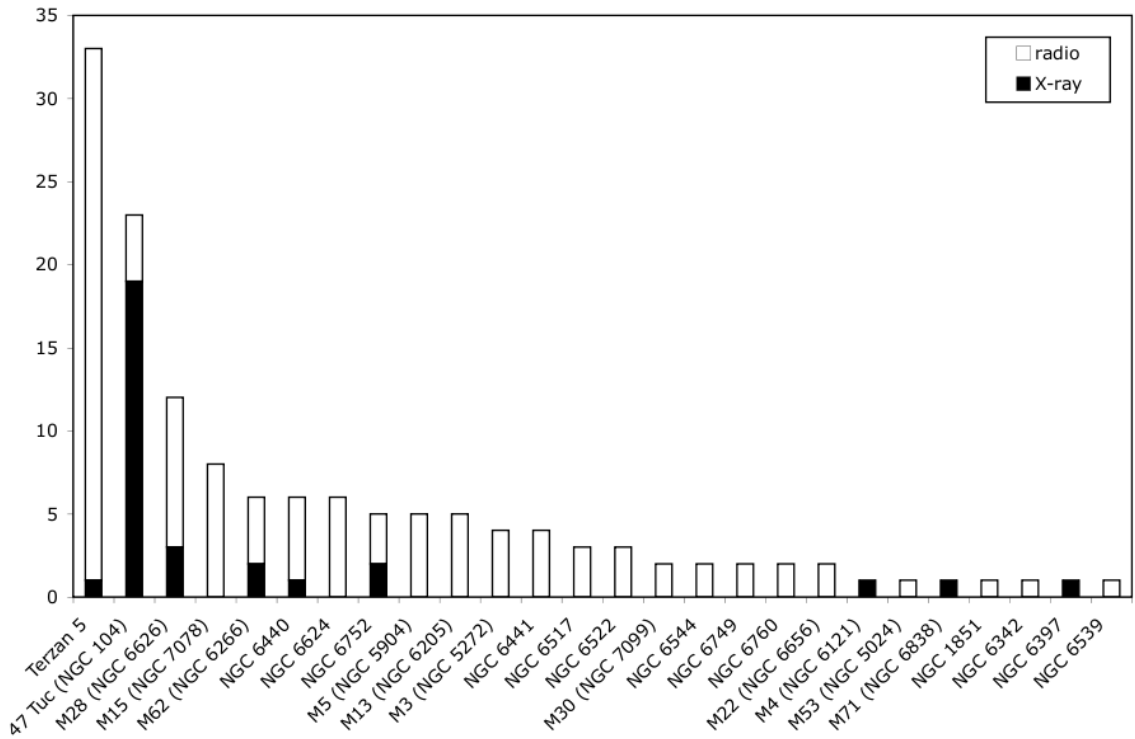,width=16cm,clip=}}
\caption[]{A histogram of radio and X-ray detected millisecond pulsars 
 in globular clusters. (Status: spring 2010).}
\label{figure1}
\end{figure}

In recent years it became obvious that globular clusters (GCs) are millisecond pulsar 
factories. Their high stellar density along with a frequent dynamical 
interaction of cluster constituents provides an efficient environment for the formation 
of short-period (binary) pulsars \citep{Rasio2000, Fregeau2008, Ivanova2008}. Extensive 
surveys using telescopes with increasing sensitivity kept the radio population 
of cluster pulsars persistently growing \citep{CamiloRasio2005, Ranson2008}. As of
spring 2010 about 160 ($\sim 9\%$) of the 1864 cataloged radio pulsars fall under the 
category of millisecond pulsars, i.e.~are recycled \citet{Manchester2005}. 140 ($\sim87\%$) 
of them are located in 26 GCs\footnote{http://www.naic.edu/$\sim$pfreire/GCpsr.html}. 
Of these globular cluster millisecond pulsars 59 ($\sim 42\%$) appear to be solitary, the 
others are in binaries. The formation scenario of solitary 
recycled pulsars is still under discussion, but it is widely believed that the pulsar's 
companion either gets evaporated (a process which is believed to be at work e.g.~in the 
PSR~B1957+20 millisecond pulsar/binary system) or that the system gets tidally disrupted 
after the formation of the millisecond pulsar. Whatever the mechanism is, it is interesting 
to note that the ratio of solitary to binary millisecond pulsars in globular clusters is 
almost identical to the 40\% observed in the population of galactic disk millisecond 
pulsars (cf.~Becker 2009). Figure~\ref{figure1} illustrates the cataloged distribution 
of X-ray and radio millisecond pulsars in each of the 26 globular clusters known to 
host millisecond pulsars.

The first millisecond pulsar found in a globular cluster was PSR~J1824-2452A (Lyne et al.~1987)
which is located in NGC 6626 (M28). PSR~J1824-2452A is not only the youngest ($P/2\dot{P}=3.0
\times 10^7$ yrs) but also the most powerful ($\dot{E}=2.24\times 10^{36} I_{45}$ erg s$^{-1}$) 
and brightest X-ray pulsar in a globular cluster. X-ray emission from it was first detected in 
ROSAT HRI data (Danner et al.~1997). With an angular resolution of $\sim 5"$ (HEW) the HRI, 
though, did not fully allow to resolve its emission into a point source. It therefore was unclear 
whether the pulsar powers a plerion or its X-ray emission was partly due to a superposition of 
multiple discrete cluster sources surrounding it. Today we know that sub-arcsecond spatial 
resolution is required to adequately study globular cluster sources in the X-ray band. Chandra 
resolved PSR~J1824-2452A into a point source along with 12 other X-ray sources surrounding it 
within the cluster's  $0^{\,\,\prime}\!.24$ core radius \citep{Becker2003}. Apparently they 
are the cause of the diffuse X-ray emission seen to surround PSR~J1824-2452A in the ROSAT HRI data.  

PSR~J1824-2452A is a strong non-thermal X-ray emitter (cf.~Becker et al.~2003, see also 
Table 6.7 in Becker 2009 for a summary of spectral parameters from X-ray detected millisecond 
pulsars). It has a relatively hard X-ray spectrum, so that X-ray pulses could be detected 
up to $\sim 20$ keV with the Rossi X-ray Timing Explorer \citep{Mineo2004}. The X-ray pulse 
profile is characterized by narrow peaks and small features with a high fraction of pulsed 
photons (Saito et al.~1997; Becker \& Tr\"umper 1999; Rutledge et al.~2004). As thermal emission 
can not be beamed strong enough to explain the narrow pulse peaks and high pulsed fraction, 
both the spectral and temporal emission properties are in agreement with the interpretation 
that its X-ray emission is caused by non-thermal radiation processes in the pulsar's magnetosphere. 

This is similar to what is observed in some other galactic plane millisecond pulsars (Becker 2009; 
Zavlin 2007; Becker \& Tr\"umper 1999). However, the data quality available from all those 
pulsars is far from being homogenous. Whereas from several of these sources high quality spectral, 
temporal and spatial information is available, for many others, especially those in globular 
clusters, the photon statistics is small. This is especially true for most of the pulsars seen 
in 47~Tuc. For this cluster Grindlay et al.~(2001) reported the detection of 108 sources 
within a region corresponding to about 5 times the 47~Tuc core radius ($r_{core}\sim 115$ arcsec). 
Nineteen of the soft/faint sources were found to be positionally coincident with millisecond 
radio pulsars. Most turned out to have a soft X-ray spectrum (Bogdanov et al.~2006) and an 
X-ray luminosity in the range $L_{X}\sim 10^{30}-10^{31}$ ergs~s$^{-1}$, which is about two
to three orders of magnitudes smaller than that of PSR~J1824-2452A.  According to Grindlay et 
al.~(2001) more than fifty percent of the unidentified sources in 47~Tuc could be millisecond 
pulsars.

Gamma-ray emission from millisecond pulsars was detected by the Fermi Large Area Telescope recently 
(Abdo et al.~2009a,b). Because of the large population of millisecond pulsars in globular 
clusters eight of them have been detected meanwhile as steady point-like high-energy gamma-ray 
sources (Abdo et al.~2010), including 47 Tuc (Webb \& Kn\"odlseder 2010).

As demonstrated by the examples above, the brightness and hardness of globular cluster pulsars 
spans a very wide range. Of the 140 cluster millisecond pulsars known today only few are found 
to have an X-ray counterpart.  To increase this sample and to study their emission properties
more systematically is the motivation for this paper. The structure is as follows: in \S2 we 
describe the source detection and identification along with the spectral and timing analysis 
of the detected millisecond pulsar counterparts. Their emission properties compared 
to those observed in Galactic plane millisecond pulsars are discussed in \S3.

\section{OBSERVATIONS AND DATA REDUCTION\label{obs}}
A decade after the launch of Chandra 18 out of the 26 globular clusters which are known
to host millisecond pulsars were observed with the ACIS and/or HRC detector in focus. 
We made use of all data publicly available by spring 2010. For counterpart searches and/or
spectral analysis there are 45 ACIS-S/I and 5 HRC-I observations available, summing up 
to a total of 1.566,419 ksec and 52.442 ksec of good data. HRC-S data with the timing-option 
enabled are available for 47 Tuc and M28, summing up to a total of 792,620 ksec and 90,440 
ksec, respectively. The corresponding observational details of these datasets are summarized 
in Table~\ref{t:observations}. Distance, half-mass radius and column density of the globular 
clusters which are known to host millisecond pulsars are summarized in Table \ref{t:clusterParameter}. 
The basic pulsar parameters for all millisecond pulsars considered in this work are summarized in
Tables~\ref{t:ms-detected-basicproperties} and \ref{t:ms-not-detected-basicproperties}.
Although several of them have been previously analyzed by various authors, leading to proposed 
counterpart assignments for some of the millisecond pulsars, new and so far unpublished data 
were available for NGC 6626 (M28), NGC 6266 (M4) and NGC 6752. 

Standard processed level-2 data were used in all counterpart searches. Prior to the data 
analysis, aspect offsets for each observation, which is a function of the spacecraft roll 
angle, was carefully checked and corrected. The data analysis was restricted to the 
$0.3-8$~keV energy range.  

\subsection{SOURCE DETECTION AND IDENTIFICATION}

To search for X-ray counterparts in the available datasets we used the WAVELET and 
CELDETECT algorithms as implemented in the CIAO 4.2 software package. Source positions 
were then correlated with the radio timing positions of the globular cluster millisecond 
pulsars. To assess all identifications we computed the probability for a chance coincidence 
according to $P_{coin}={ (N^{\rm all}_{X}/{r_{\rm s}^{2})~\delta{\rm RA}~\delta{\rm Dec}}}\,$.
Here $N^{\rm all}_{X}$ is the number of X-ray sources detected within a search region of
radius $r_{\rm s}$ and  $\delta{\rm RA}$ $\delta{\rm Dec}$ are the positional uncertainties 
in right ascension and declination. The latter was determined by combining the errors of the 
radio pulsar and X-ray source positions in quadrature, considering Chandra's absolute 
astrometric accuracy to be $0.21''$. This astrometric accuracy was estimated from the 
distribution of aspect offsets for a sample of point sources with accurately known celestial 
positions\footnote{See http://cxc.harvard.edu/cal/ASPECT/celmon/ for details}. There are 
$68\%$ of 237 sources imaged with ACIS-S which have offsets smaller than or equal to 
$\sim0.21''$. As all the data used for counterpart searches were taken with 
the ACIS-S we included this value in our error budget computation for each coordinate. 
For the search radius $r_{\rm s}$ we used the globular cluster half-mass radius as most 
of the pulsars lie within this region. The pulsar PSR~J$1911-6000$C in NGC~6752, though, 
is located outside this radius so that we expanded the search region for this cluster 
to 3 arcmin. The resulting identifications, counting rates and chance probabilities for 
finding an X-ray source by chance at the radio pulsar position are listed in 
Table~\ref{t:xraydetections}.

In total 31 X-ray sources in nine globular clusters were found to be coincident with 
the radio timing position of known millisecond pulsars. Among them are the 19 
millisecond pulsars previously identified 
in 47~Tuc \citep{Grindlay2001, Bogdanov2006}, the millisecond pulsar PSR~J1824-2452A in 
NGC 6626 \citep{Becker2003}, PSR~1953+1846A in NGC 6838 \citep{Elsner2008}, PSR~J1701-3006B 
in NGC 6266 \citep{Cocozza2008}, PSR~J1740-5340 in NGC 6397 \citep{HunangBecker2010} and few 
more for which an association has been assigned in the literature (cf.~Table 6.9 in Becker 
2009 and references therein). Pulsars for which X-ray counterparts are newly detected are 
PSR~J1824-2452G and PSR~J1824-2452H in NGC 6626 and PSR~J1701-3006C in NGC 6266. The tentative 
assignment of the X-ray counterparts for PSR~J2140-3310A in M30 \citep{Ransom2004} and 
PSR~J1910-5959B \citep{DAmico2002} could not be confirmed by our analysis, albeit additional 
data were available for the latter pulsar compared to their analysis. 

\begin{figure}
\centerline{\psfig{figure=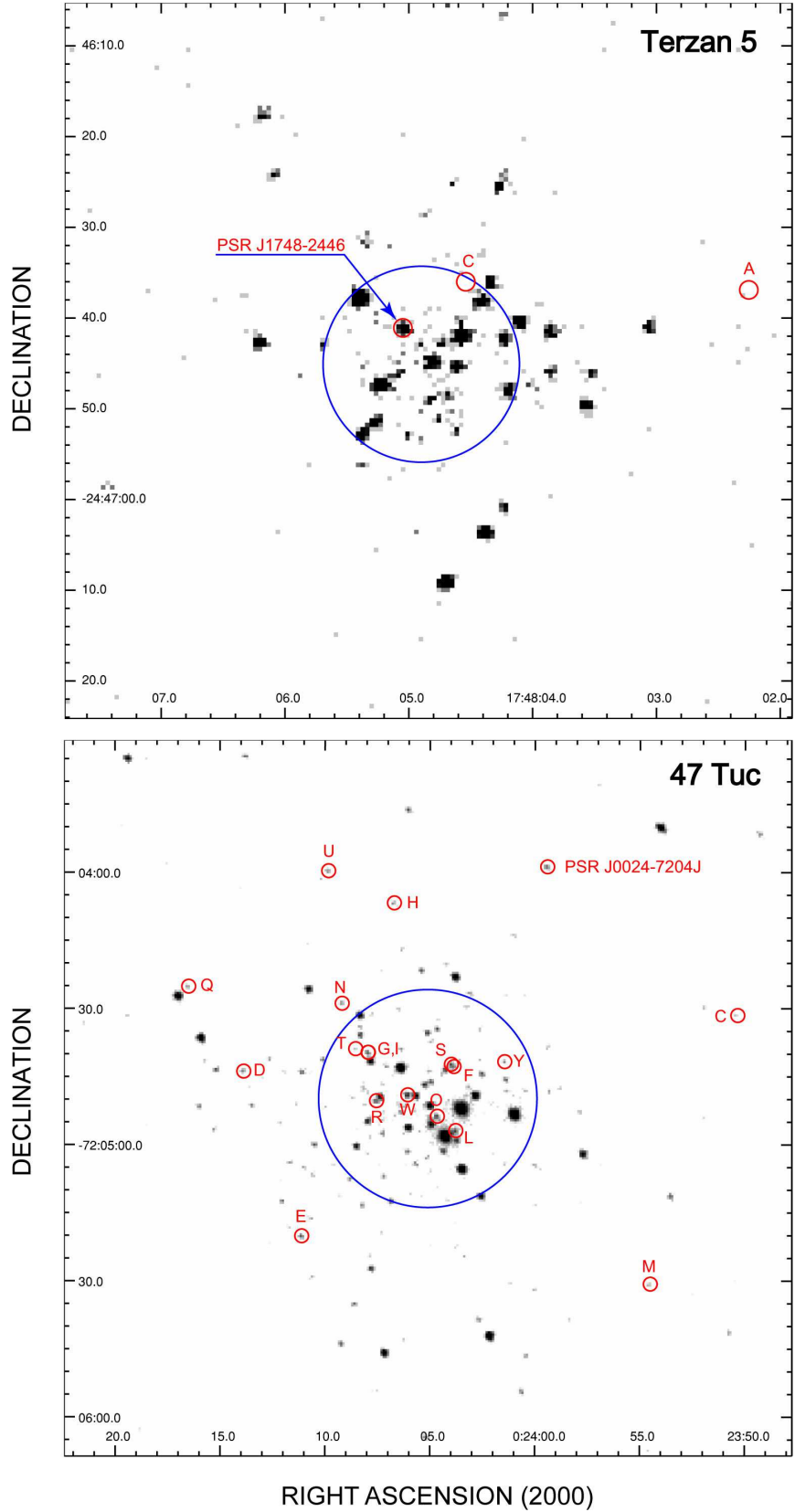,height=22cm,clip=}}
\caption[]{Chandra ACIS-S view on the globular clusters Terzan 5 and 47 Tuc. 
The clusters core radius and the position of known millisecond pulsars 
are indicated by blue and red circles, respectively. (Status: spring 2010).}
\label{figure2}
\end{figure}

\begin{figure}
\centerline{\psfig{figure=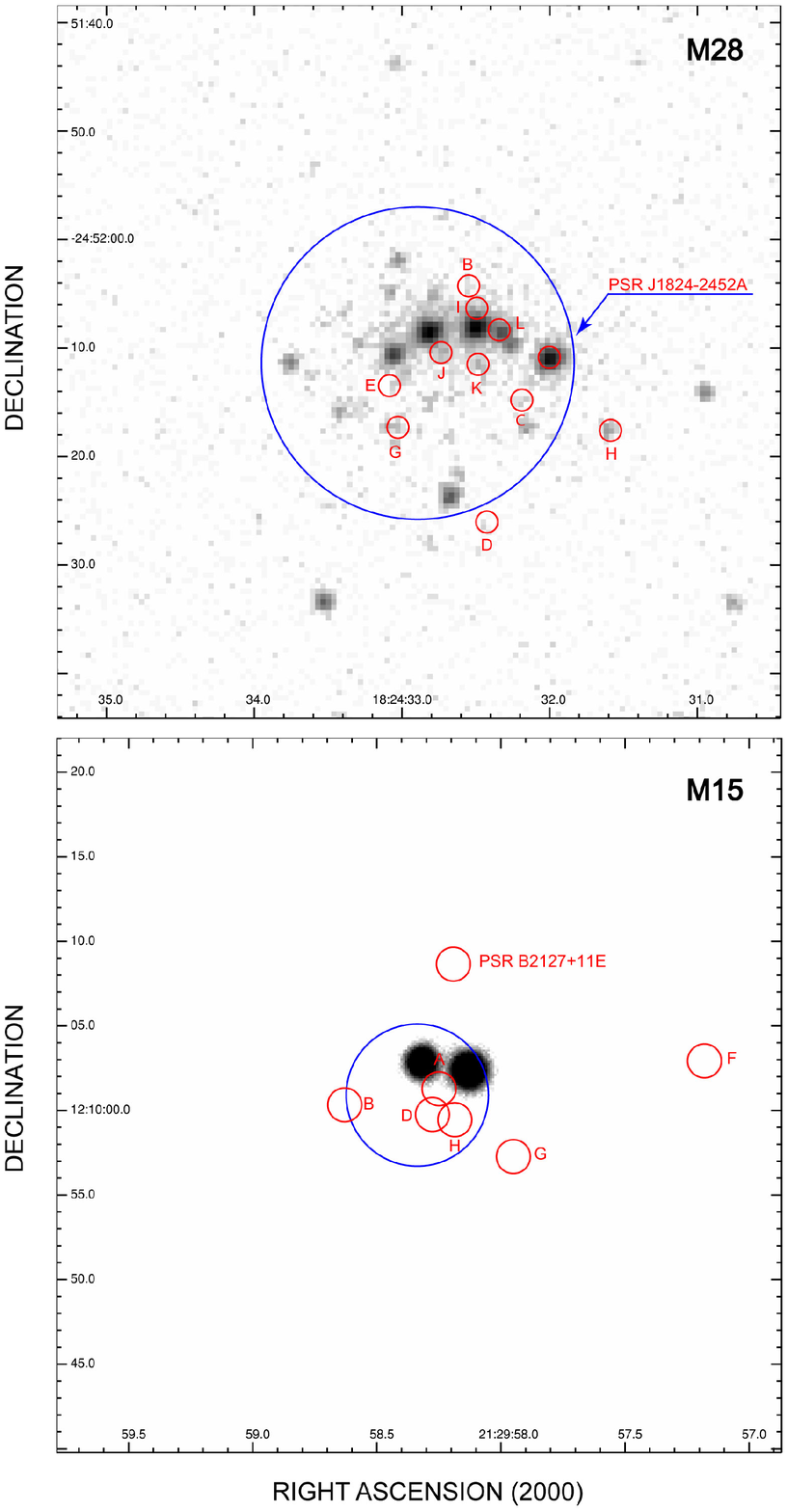,height=22cm,clip=}}
\caption[]{Chandra ACIS-S view on the globular clusters M28 (NGC 6626) 
and the HRC-I view on M15 (NGC 7078). The clusters core radius and 
the position of known millisecond pulsars are indicated by blue and red 
circles, respectively. (Status: spring 2010).}
\label{figure3}
\end{figure}

\begin{figure}
\centerline{\psfig{figure=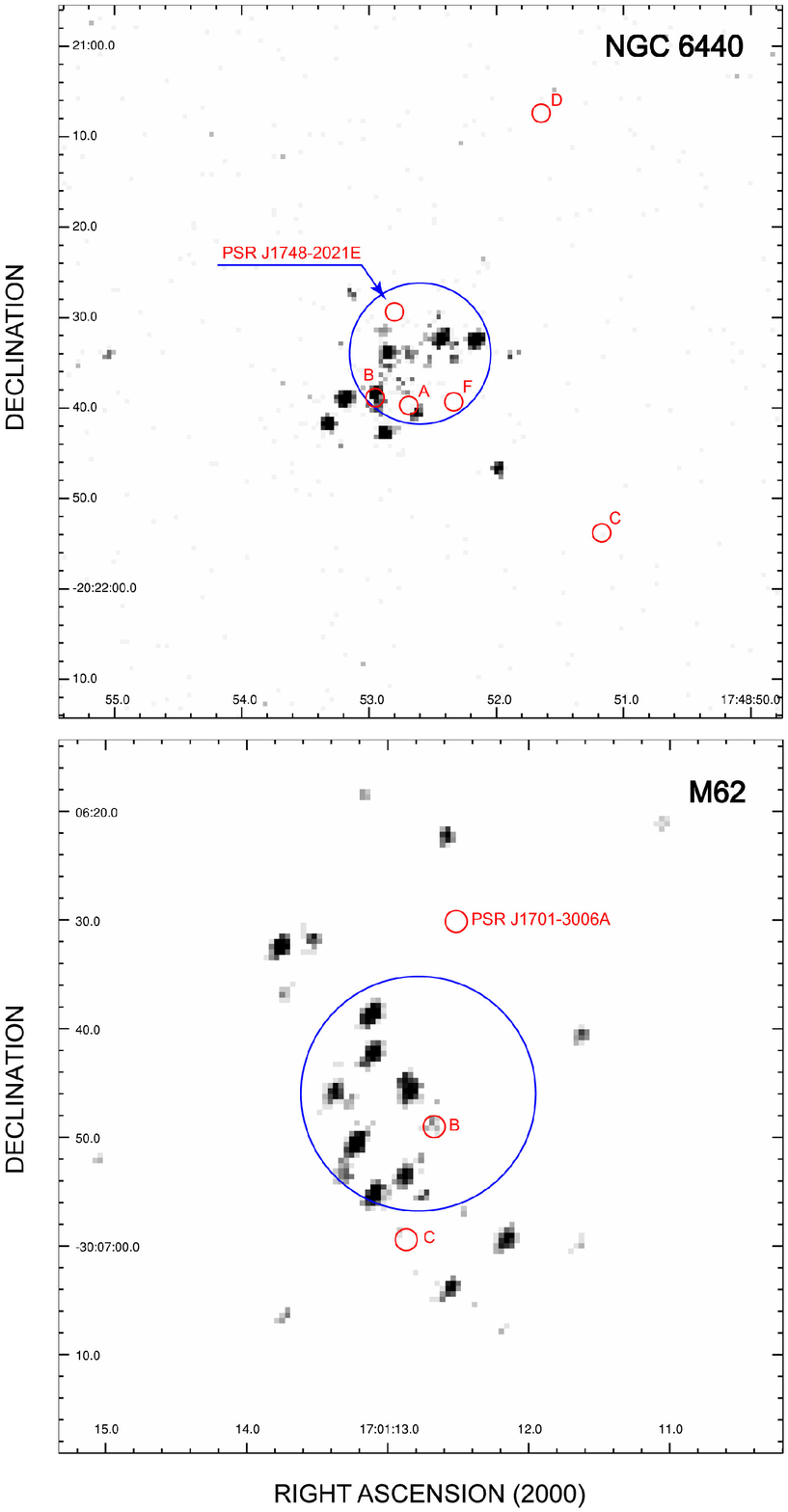,height=22cm,clip=}}
\caption[]{Chandra ACIS-S view on the globular clusters NGC 6440 and
M62 (NGC 6266). The clusters core radius and the position of known
millisecond pulsars are indicated by blue and red circles, respectively.
(Status: spring 2010).}
\label{figure4}
\end{figure}

\begin{figure}
\centerline{\psfig{figure=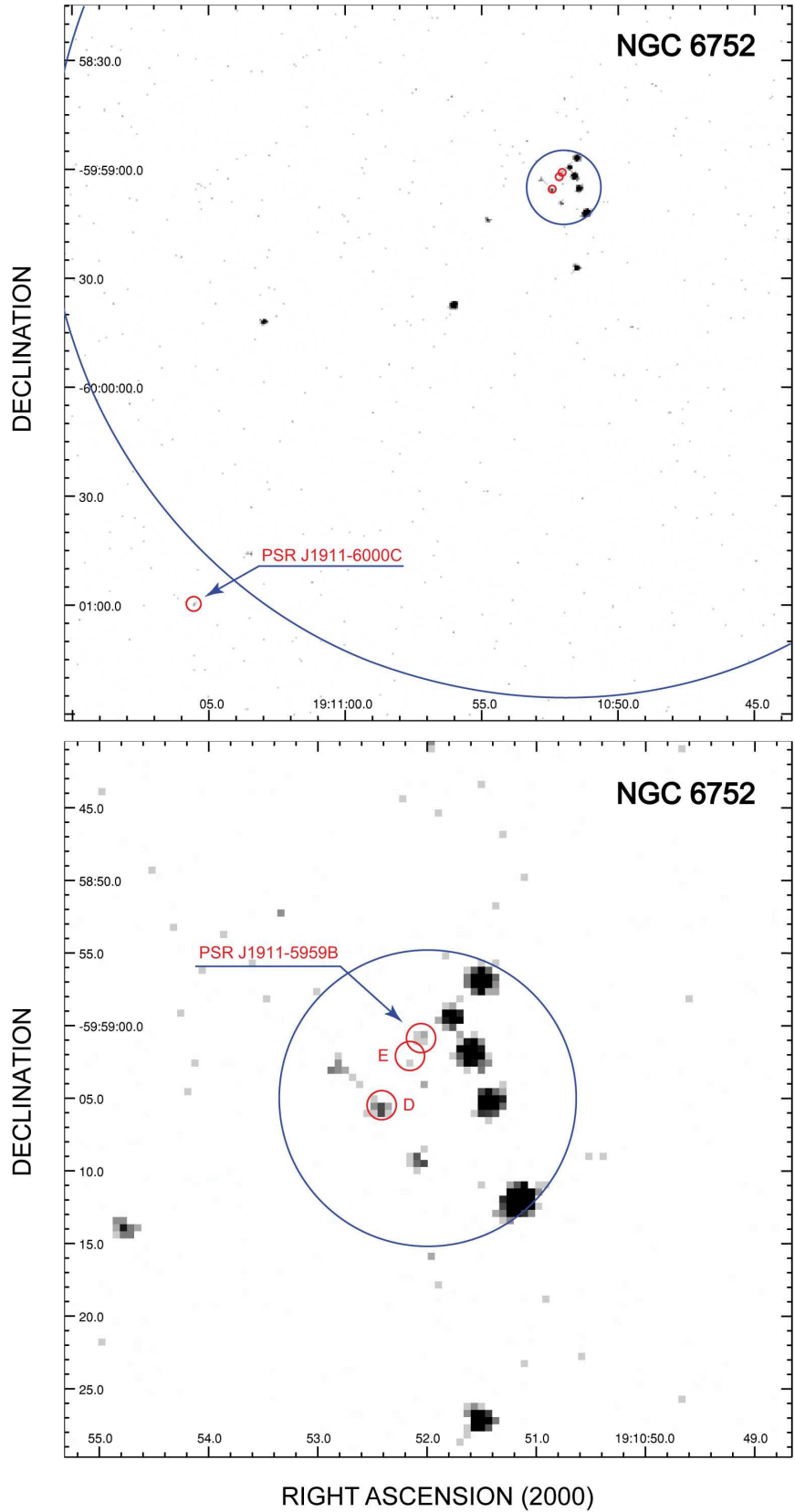,height=22cm,clip=}}
\caption[]{Chandra ACIS-S view on the globular clusters NGC 6752. The 
clusters core and half-mass radius (inner/outer circles) and the 
position of known millisecond pulsars are indicated by blue and red 
circles, respectively. The bottom figure shows a zoom in of the top 
figure. (Status: spring 2010).}
\label{figure5}
\end{figure}

\begin{figure}
\centerline{\psfig{figure=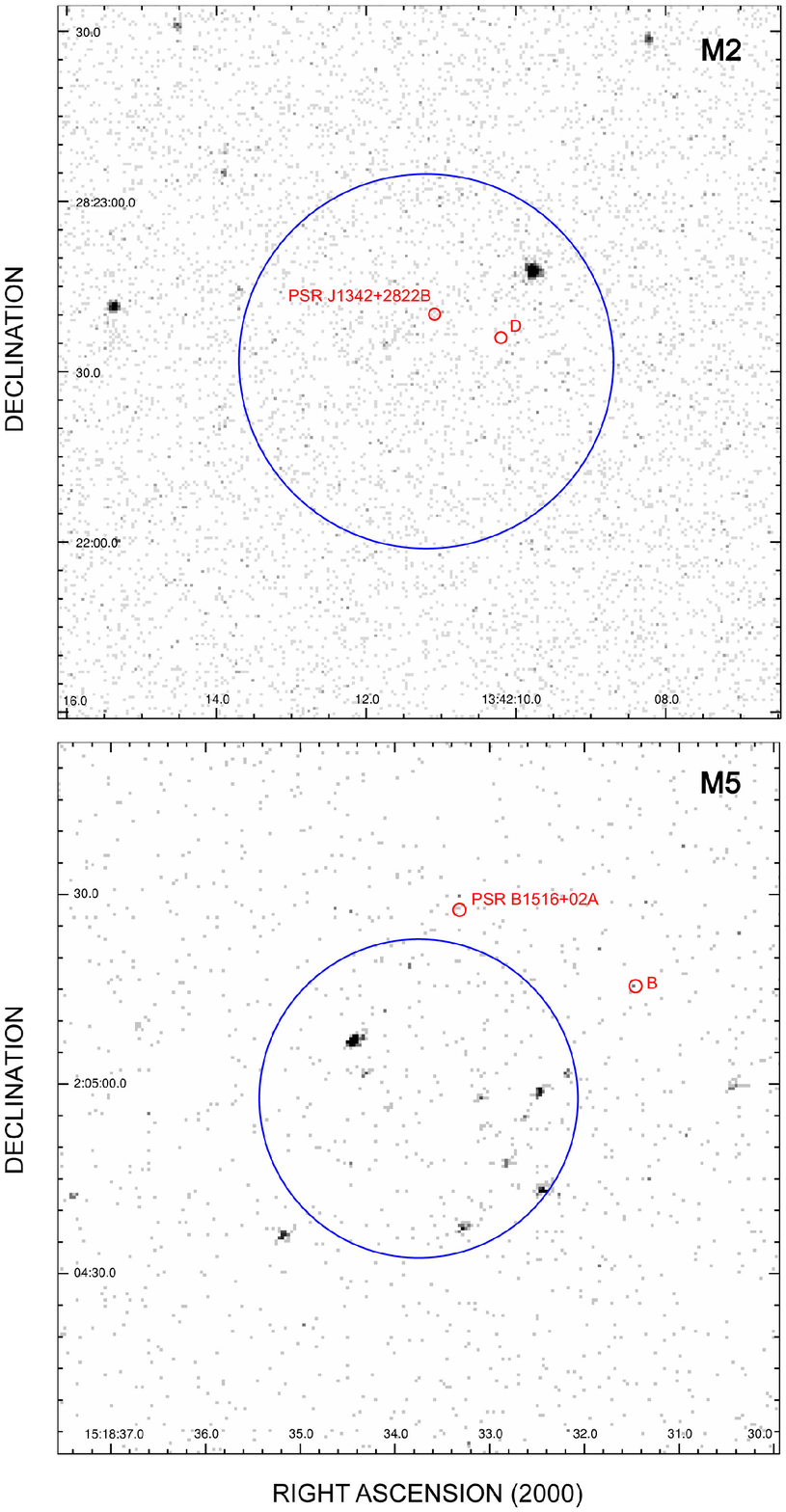,height=22cm,clip=}}
\caption[]{Chandra ACIS-S view on the globular clusters M3 (NGC 5272) and
M5 (NGC 5904). The clusters core radius and the position of known 
millisecond pulsars are indicated by blue and red circles, respectively. 
(Status: spring 2010).}
\label{figure6}
\end{figure}

\begin{figure}
\centerline{\psfig{figure=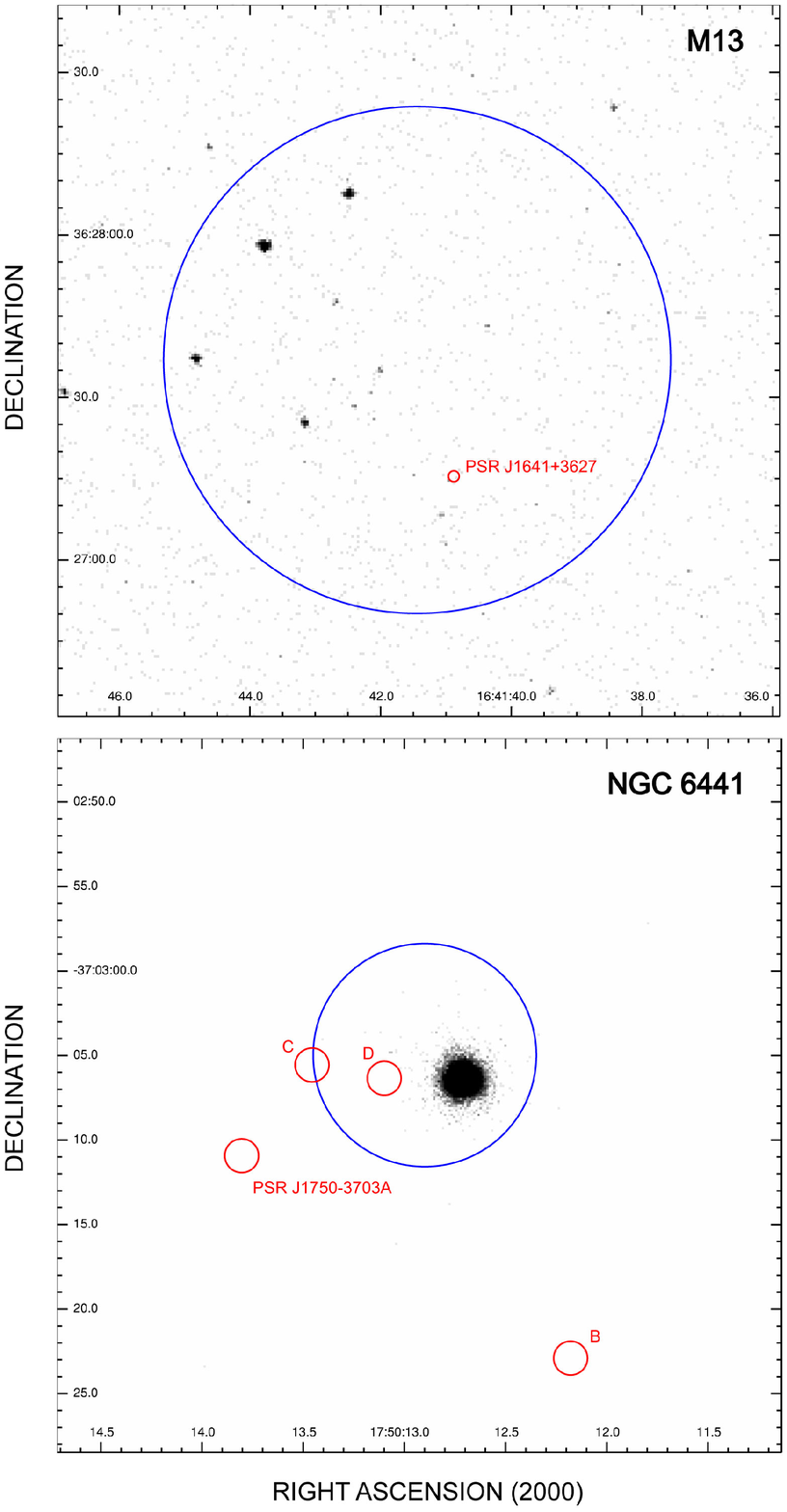,height=22cm,clip=}}
\caption[]{Chandra ACIS-S view on the globular clusters M13 (NGC 6205) and
the HRC-I view on NGC 6441. The clusters core radius and the position 
of known millisecond pulsars are indicated by blue and red circles, respectively.
(Status: spring 2010).}
\label{figure7}
\end{figure}

\begin{figure}
\centerline{\psfig{figure=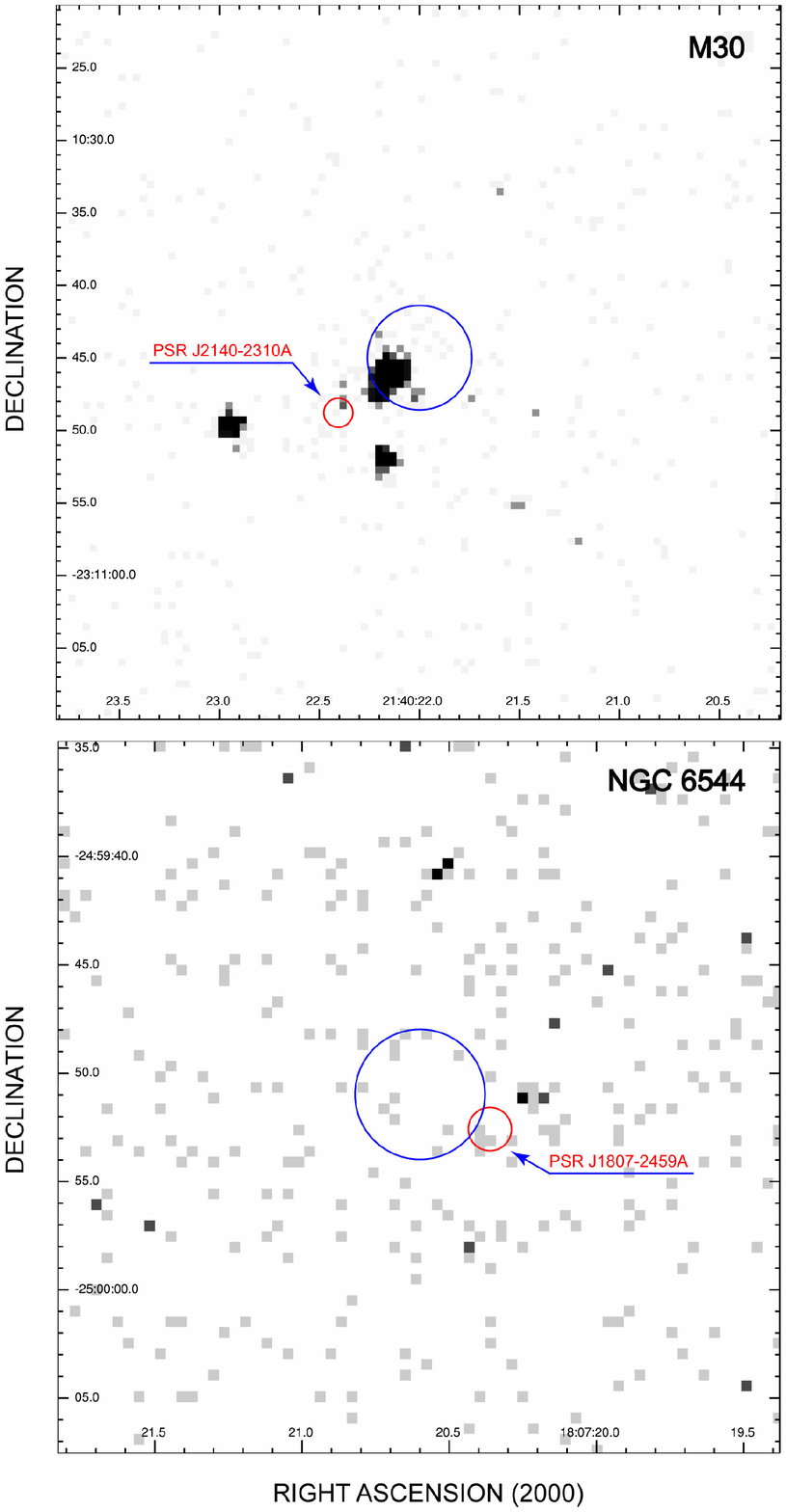,height=22cm,clip=}}
\caption[]{Chandra ACIS-S view on the globular clusters M30 (NGC 7099) and
NGC 6544. The clusters core radius and the position of known
millisecond pulsars are indicated by blue and red circles, respectively.
(Status: spring 2010).}
\label{figure8}
\end{figure}

\begin{figure}
\centerline{\psfig{figure=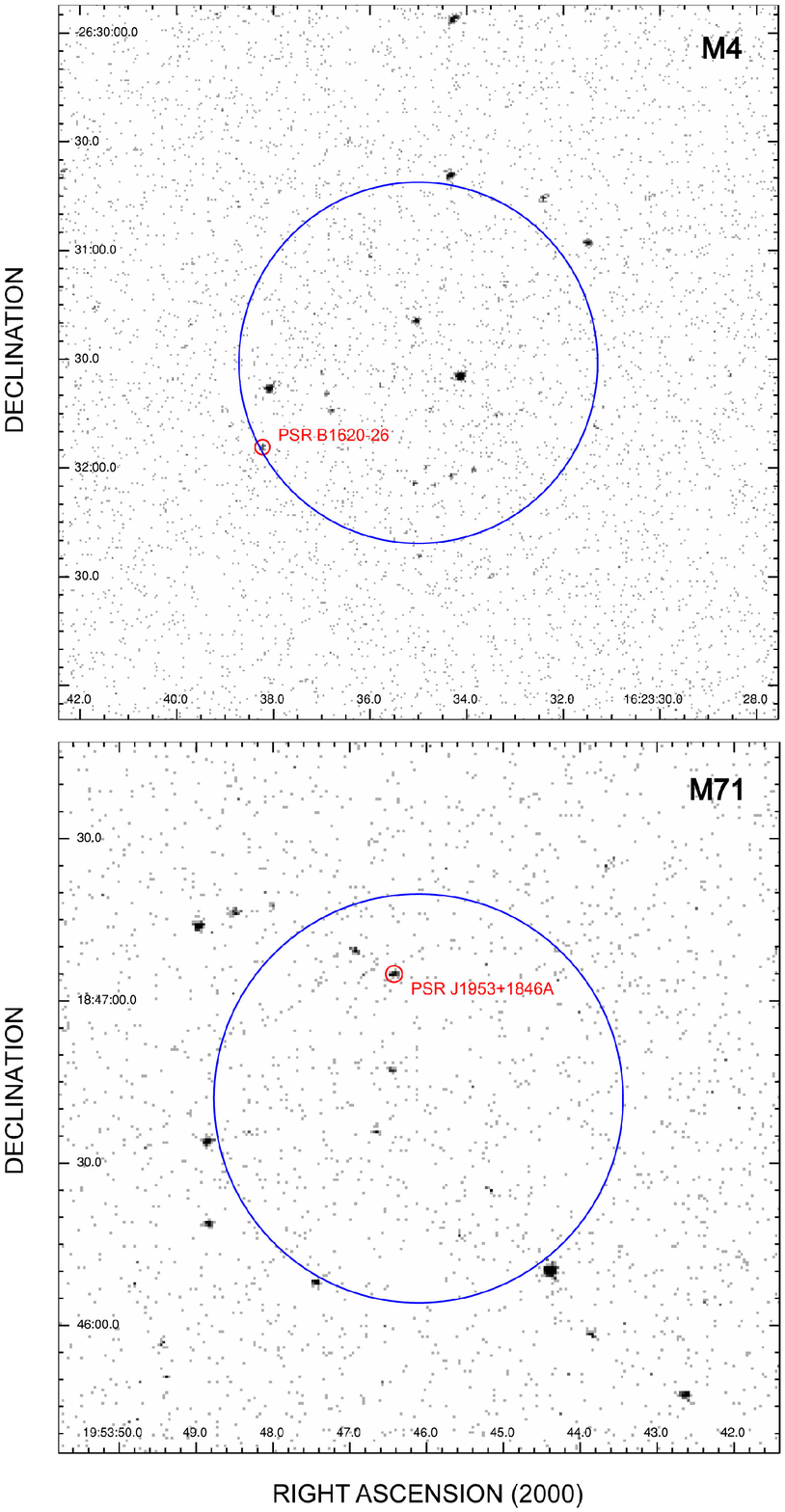,height=22cm,clip=}}
\caption[]{Chandra ACIS-S view on the globular clusters M4 (NGC 6121) and
M71 (NGC 6838). The clusters core radius and the position of known
millisecond pulsars are indicated by blue and red circles, respectively.
(Status: spring 2010).}
\label{figure9}
\end{figure}

\begin{figure}
\centerline{\psfig{figure=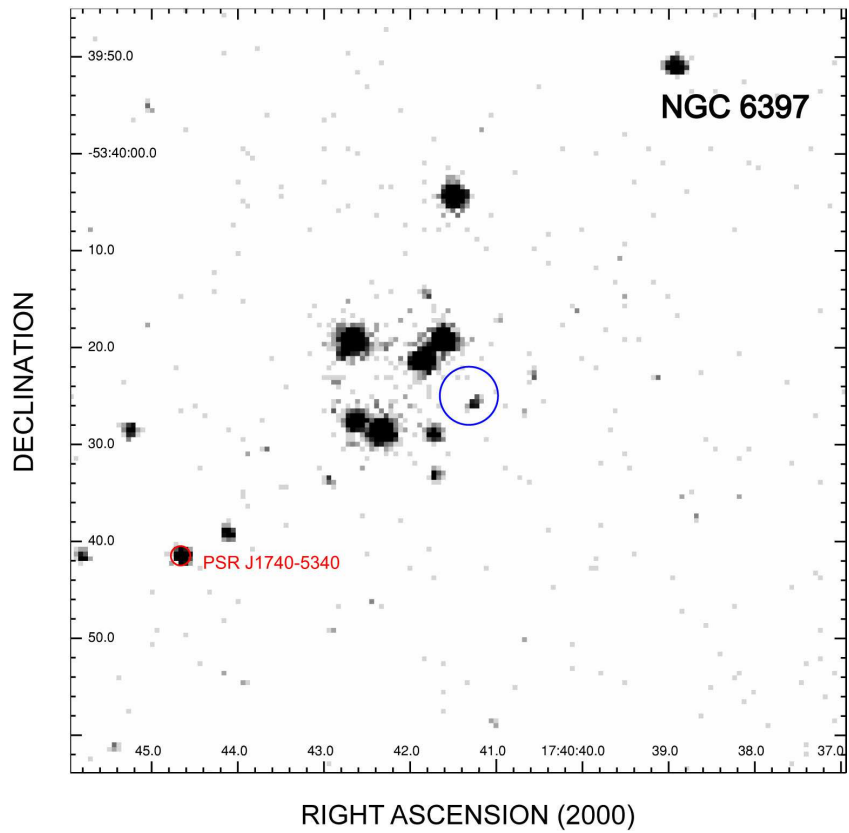,height=11cm,clip=}}
\caption[]{Chandra ACIS-S view on the globular clusters NGC 6397. 
The clusters core radius and the position of known millisecond 
pulsars are indicated by blue and red circles, respectively.
(Status: spring 2010).}
\label{figure10}
\end{figure}

Counting rate upper limits were computed for those pulsars for which no X-ray counterpart
could be detected. For this we measured the number of counts recorded at the radio pulsar 
position and computed the $3\sigma$ upper limits according to $C_{3\sigma}=0.5 \times 
(S/N)^2+(S/N) \times \sqrt{cts+0.25 \times (S/N)^{2}}$. Here $S/N=3$ is the signal-to-noise 
ratio and $cts$ the counts obtained within a circle of 1 arcsec radius centered on the position
of the radio pulsar. All upper limits are summarized in Table \ref{t:upperlimits} along with 
the exposure time of the observation and the number of counts recorded at the pulsar position. 
Figures \ref{figure2} to \ref{figure10} show ACIS-S and/or HRC-I images of all globular clusters 
considered in this work. The location of the millisecond pulsars and the cluster half-mass radius
are indicated.

\subsection{TIMING ANALYSIS AND SEARCHES FOR LONG TERM FLUX VARIABILITY}

\subsubsection{THE GLOBULAR CLUSTER X-RAY PULSARS}

For about ten years the 3.05 ms pulsar PSR~J1824-2452A in M28 was the only globular cluster 
pulsar from which pulsed X-ray emission had been detected. Timing observations were performed 
with ROSAT (Danner et al.~1997), ASCA (Saito et al.~1997), BeppoSax (Mineo et al.~2004) and 
the Rossi X-ray Timing Explorer (e.g.~Rots 2006). XMM-Newton lists the pulsar as calibration
target, although it was never scheduled for observations so far. Chandra observed
PSR~J1824-2452A in 2002 and 2006 for a total of $\sim 90,440$ ksec using the HRC-S with the 
timing flag enabled. The temporal resolution of the HRC-S in this mode is $15.625\,\mu s$. 
It is sensitive in the $0.08-10$ keV band\footnote{http://hea-www.harvard.edu/HRC/overview/overview.html}. 
The intrinsic energy resolution of the HRC being a channel plate detector is marginal.
Rutledge et al.~(2004) used the 2002 data together with a series of archival XTE data to 
re-investigate the pulsars temporal X-ray emission properties. With respect to their analysis
we included the HRC-S observation taken in 2006 which almost doubles the number of photons 
available for the Chandra timing analysis of PSR~J1824-2452A. This allowed us to measure 
the pulse arrival time, fraction of pulsed photons, the width of the two X-ray pulses and 
the pulse phase separation of the two peaks by using the Chandra data alone. As Chandra is 
an imaging instrument it allows to measure the pulse properties by its better 
signal-to-noise ratio with much higher accuracy than using XTE data only. 

\begin{figure}
\centerline{\psfig{figure=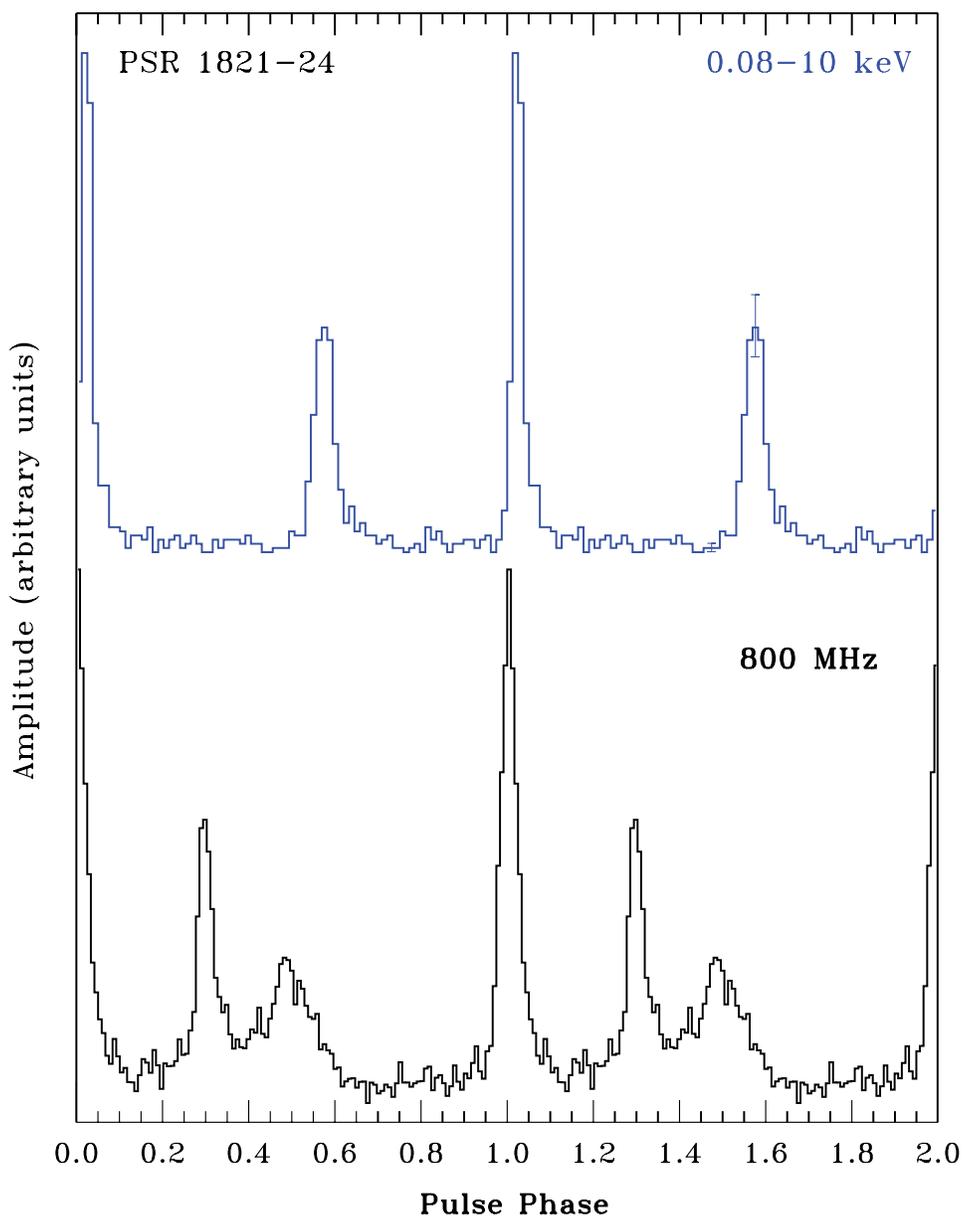,height=16cm,clip=}}
\caption[]{Integrated pulse profiles of the globular cluster pulsar PSR J1824$-$2452A
  as observed with the Chandra HRC-S ({\em top}) and with the NRAO at 800 MHz ({\em bottom}) 
  by Backer \& Sallmen (1997). Two phase cycles are shown for clarity. The X-ray pulse 
  profiles are characterized by two narrow peaks with a phase separation of 
  $\sim 0.449\pm 0.0009$ between the two peaks. The radio profile at 800 MHz depicts 
  three pulse components, with the main radio peak leading the second and third. The 
  dominating radio pulse is leading the main X-ray pulse by $0.0243$ in phase. The 
  uncertainty of the ralative phase is $\pm 1$ bin ($\sim 38 \mu\mbox{s}$) in the 
  X-ray profile. Phase zero corresponds to JD(TDB@SSB)=2451468.5009909072.}
\label{figure11}
\end{figure}

Selecting all events from within a circle of 1.5 arcsec radius centered on the pulsar position
gave us a total of 773 counts to construct a pulse profile by period folding (Buccheri \& De Jager 
1989). The background contribution is estimated to be $45 \pm 7$ counts. The pulsar ephemeris which 
were used to fold the arrival times with the pulsar's rotation period P=3.0543151 ms, predicted for 
the epoch JD=2451468.5, are summarized in Table \ref{t:ephemeris}. The X-ray pulse profile relative 
to the radio profile taken at 800 MHz by Backer \& Sallmen (1997) is shown in Figure \ref{figure11}. 
Unlike the radio profile, which is observed to show three peaks of which the second is observed to 
vary in intensity on time scales of weeks to month (Backer \& Sallmen 1997), only two narrow pulse 
peaks are observed in the X-ray band. Their phase separation obtained from fitting the pulse 
peaks by Gaussian functions is found to be $0.449 \pm 0.0009$. The FWHM of the phase width of the 
first and second peaks are fitted to be $0.0273\pm 0.0009$ and $0.053 \pm 0.002$, corresponding to 
$9.8\pm 0.3$ and $19.3\pm 0.7$ degrees. The three radio pulse peaks have a phase separation of 
$\mbox{RP2}-\mbox{RP1}= 0.295 \pm 0.001$ and $\mbox{RP3}-\mbox{RP2}= 0.193 \pm 0.003$. The FWHM
phase width of the radio pulse peaks are fitted to be $0.0452 \pm 0.0013$, $0.037 \pm 0.002$ and 
$0.16 \pm 0.01$ for the main, second and third radio pulse. The X-ray pulsed fraction is determined 
to be $85\pm 4\%$ by using a bootstrap method (cf.~Becker \& Tr\"umper 1999; Swanepoel et al.~1996). 
The error represents the $1\sigma$ confidence range. For the phase difference between the radio and X-ray 
peak it can be seen from Figure \ref{figure11} that the main radio pulse leads the main X-ray 
pulse by $0.0243\pm0.0004$ in phase, corresponding to $74.2 \pm 1.2\, \mu\mbox{s}$. Rots (2006) has 
recently estimated the accuracy of Chandra's absolute time stamps to be $4\pm 4\,\mu\mbox{s}$. 
A $3\sigma$ clock uncertainty of $16\,\mu\mbox{s}$ together with the HRC's intrinsic temporal resolution 
thus results in a total uncertainty of $32\,\mu\mbox{s}$ for the absolute phase assigned to an X-ray 
photon. Using 79 phase bins to construct the X-ray profile one bin corresponds to $\sim 38\,\mu\mbox{s}$ 
of the pulsar's rotation period. The uncertainty of the relative phase between the radio and X-ray 
profiles plotted in Figure \ref{figure11} thus corresponds to $\pm 1$ phase bin in the X-ray profile. 

\begin{figure}
\centerline{\psfig{figure=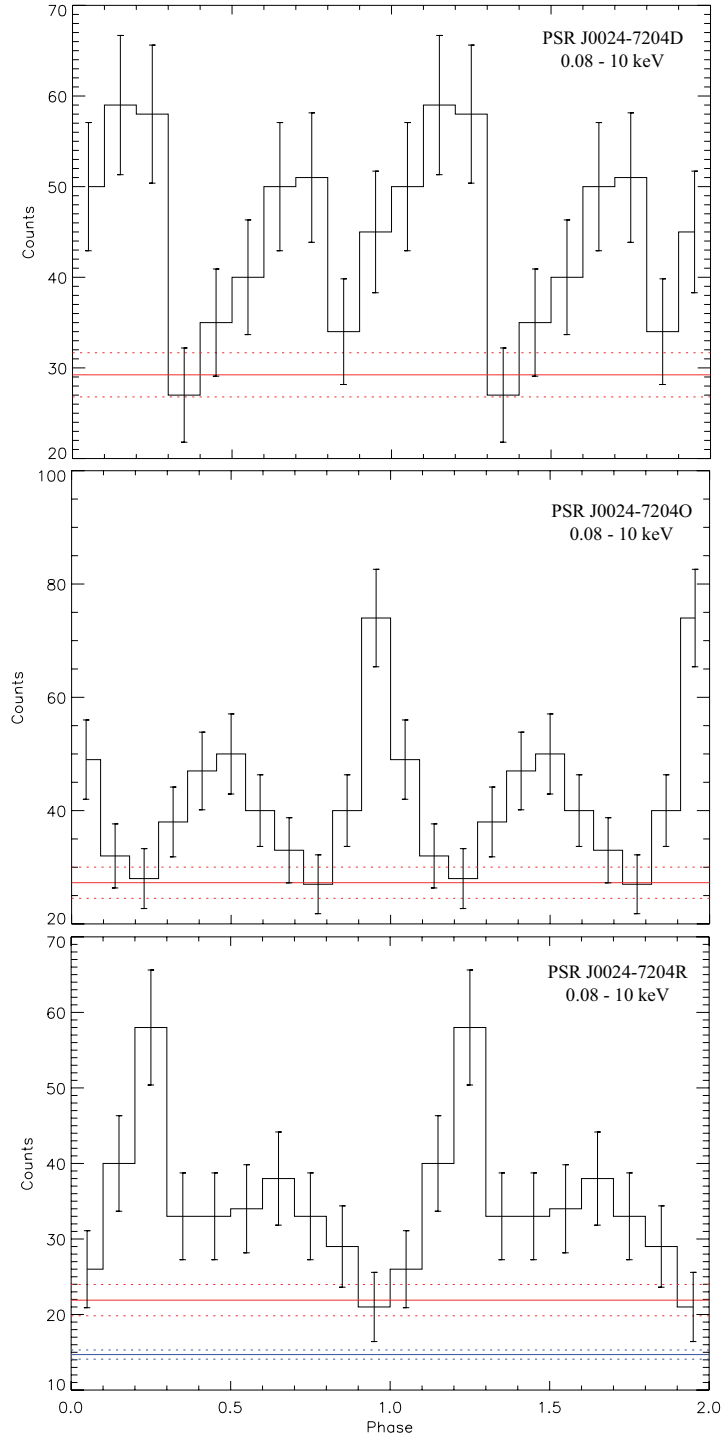,height=19cm,clip=}}
\caption[]{Integrated pulse profiles of the 47Tuc pulsars PSR J0024$-$7204\,D,O,R.
  Two phase cycles are shown for clarity. The profiles are characterized by two pulse
  peaks per rotation period. The red solid and dotted lines indicate the DC level 
  and its $1\sigma$ uncertainty. The blue solid and dotted lines in J0024$-$7204R 
  indicate the level of background contribution which is $14.7\pm 1.2$ cts/bin. 
  For J0024$-$7204D and J0024$-$7204O it is $18.6 \pm 1$ and $17\pm 1.2$ cts/bin,
  respectively. Phase zero corresponds to JD(TDB@SSB)=2451600.5, 2453733.50981, and 
  2453734.15963 in the plots of PSR J0024$-$7204\,D,O,R.}
\label{figure12}
\end{figure}

\begin{figure}
\centerline{\psfig{figure=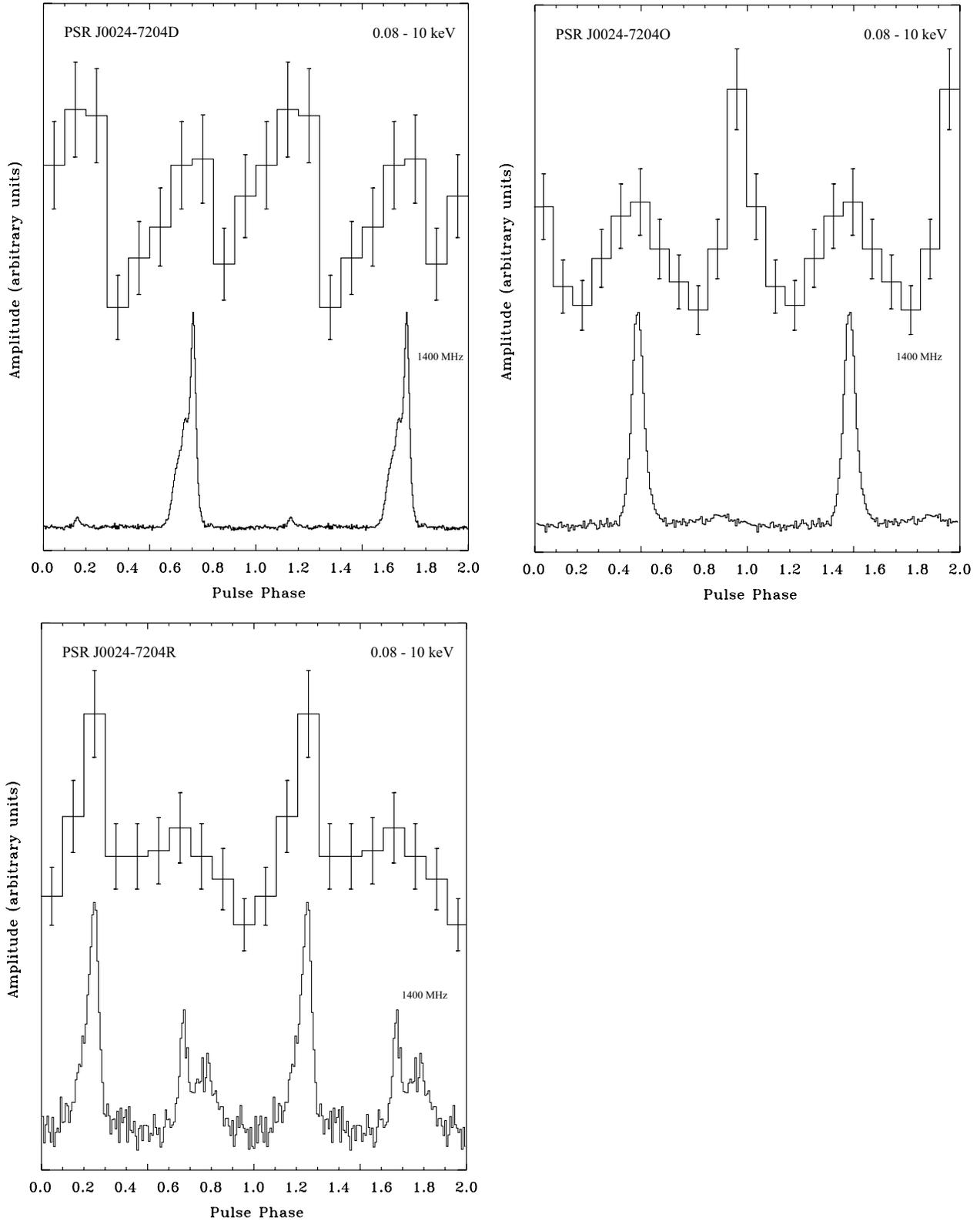,height=21cm,clip=}}
\caption[]{Integrated X-ray and radio pulse profiles of the 47Tuc pulsars PSR J0024$-$7204\,D,O,R. 
  Two phase cycles are shown for clarity. The relative phase allignement is arbitrary. The X-ray 
  and radio profiles of J0024$-$7204R show an interesting gross similarity. (Radio profiles from 
  Freire et al.~2010, im prep.)}
\label{figure13}
\end{figure}

The only other globular cluster millisecond pulsars for which pulsed X-ray emission is detected
are located in 47 Tuc. So far, $\sim 800$ ksec deep Chandra observations found X-ray pulses 
from only three of the nineteen 47 Tuc millisecond pulsars \citep{Cameron2007}. We used the archival 
HRC-S data (cf.~Table \ref{t:observations}) to re-analyze these data in order to consistently 
determine their temporal emission properties. The ephemeris we used are summarized in Table 
\ref{t:ephemeris}. X-ray pulses are reported for PSR J0024-7204D, PSR J0024-7204O and PSR J0024-7204R. 
PSR J0024-7204D is a solitary pulsar spinning at a period of 5.35 ms. PSRs J0024-7204\,O,R are in 
short-period binaries (Camilo et al.~2000) with a $\sim 0.02$ M$_\odot$ companion. Their spin\,/\,orbit 
periods are 2.64 ms\,/\,3.26 h and 3.48 ms\,/\,1.58 h, respectively. For analyzing their temporal 
emission properties we selected all events from circular regions of radius 1 arcsec or 1.5 arcsec, 
centered on the pulsar position. The smaller selection radius was used for J0024-7204\,O,R in 
order to minimize a possible flux contribution from neighboring sources located in the crowded cluster 
center (cf.~Figure \ref{figure2}). The barycentered photon arrival times were coherently folded 
with the spin-period extrapolated to the epoch $t_0$ listed in Table \ref{t:ephemeris}. The 
$z^2_n$-test (Buccheri \& De Jager 1989) was applied for $n=1-10$ harmonics in combination with 
the H-test (de Jager 1987) to determine the significance of the pulsed signal as a function of 
its harmonic content. According to this tests the pulsations in PSR J0024-7204\,D,R have the highest 
significance of $3.64\,\sigma$, and $3.96\,\sigma$ for $n=2$ harmonics whereas the pulsed signal 
in J0024-7204O is found to have a significance of $4.84\sigma$ for $n=3$ harmonics. The pulse 
profiles are shown in Figure \ref{figure12}. They all appear to be double peaked with a phase 
separation of $0.44\pm 0.06$, $0.50\pm 0.02$ and $0.4\pm 0.04$ for PSRs J0024-7204\,D,O,R. 
The phase width of the two peaks are $0.4\pm 0.1$ (FWHM) and $0.25\pm 0.15$ (FWHM) for PSR 
J0024-7204\,D, $0.15\pm 0.04$ (FWHM) and $0.27\pm 0.02$ (FWHM) for PSR J0024-7204\,O and 
$0.2\pm 0.06$ (FWHM) and $0.32\pm 0.05$ (FWHM) for PSR J0024-7204R. For the fraction of pulsed 
photons we measured $60\pm 15 \%$ for PSR J0024$-$7204\,D, $57\pm 15\%$ for PSR J0024$-$7204\,O and
$64\pm 17 \%$ for PSR J0024$-$7204\,R using a bootstrap method (cf.~Becker \& Tr\"umper 1999; Swanepoel 
et al.~1996). Figure \ref{figure13} shows the pulsar's X-ray profiles together with their corresponding 
radio profiles observed at 1400 MHz by Freire et al.~(2010, in prep.).

\subsubsection{SEARCHES FOR LONG TERM FLUX VARIABILITY}
The 3.2 s frame time of the ACIS-S detector does not allow to search for 
coherent X-ray pulsations from the clusters' millisecond pulsars. Most of the 
sources considered in this work, though, were in the focus of the ACIS-S for 
multiple times, permitting us to investigate their temporal behavior on 
longer time scales. The length of the time scales in the various datasets 
is given by the time gaps between the different observations, e.g.~hours to years.

\begin{figure}
\centerline{\psfig{figure=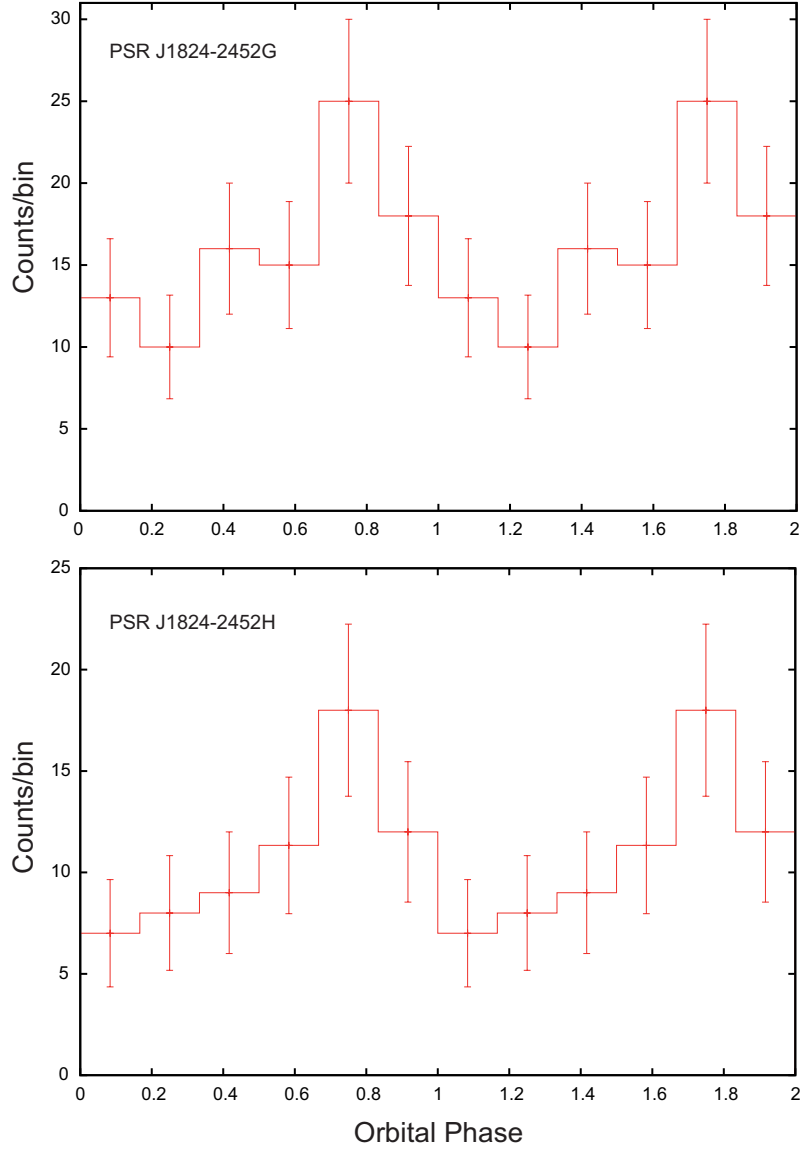,height=16cm,clip=}}
\caption[]{Lightcurves of the pulsars PSR J1824-2452G (top) and 
J1824-2452H (bottom) in M28. The data were binned into six phase 
bins. Error bars indicate the $1\sigma$ uncertainty. Two phase 
cycles are shown for clarity.}
\label{figure14}
\end{figure}

We checked the counting rates of all pulsars for variability (cf.~Table \ref{t:xraydetections}). 
Binary pulsars (cf.~Table \ref{t:ms-detected-basicproperties}) were tested 
on whether they show flux variability related to e.g.~their orbital binary 
motion. This was observed in 47 Tuc W \citep{Bogdanov2005} which exhibits 
large-amplitude X-ray variability, probably due to geometric occultations of 
an X-ray emitting intra-binary shock by the companion main-sequence star 
\citep{Bogdanov2006}.
As an example, Figure \ref{figure14} shows the lightcurves of the M28 pulsars 
PSR~J1824-2452G and J1824-2452H as folded at their binary period. PSR~J1824-2452G 
is in a 2.51\,h binary system with a low-mass companion. J1824-2452H has an orbit 
period of 10.87\,h and has been observed to show radio eclipses \citep{Begin2006}. 
Indeed, inspecting the lightcurves by eye may reveal a flux variability with the 
pulsar's orbital binary motion. From the statistical point of view, though, the
significance for variability depends on the number of phase bins and is only at 
the $\sim 1.5-2\sigma$ level for lightcurves with 6, 8 or 10 phase bins. 

More stringent results were found for PSR~J1748-2021B in NGC 6440 and for PSR~J1740-5340 
in NGC~6397. PSR~J1748-2021B has an orbital period of $\sim20$ days \citep{Freire2008}. It 
was observed by Chandra twice, in 2000 July 4 and three years later in 2003 June 27. 
These observations cover the relatively narrow orbital phase intervals\footnote{The zero point 
is set to the ascending node.} $0.124-0.137$ and $0.058-0.072$. 
The vignetting  corrected net counting rates which we have measured from the pulsar in this 
two datasets are $(1.68\pm 0.28) \times10^{-3}$ cts/s and $(5.03\pm0.46)\times10^{-3}$ cts/s, 
respectively. The counting rates measured in both data thus differ by a factor of $\sim3$ 
with a significance of $\sim 6\,\sigma$.  On whether this variability is because of a 
modulation of the X-ray flux over the pulsar's binary orbit or because of a long term 
flux increase by other means can not be clarified with the available data.

\begin{figure}
\centerline{\psfig{figure=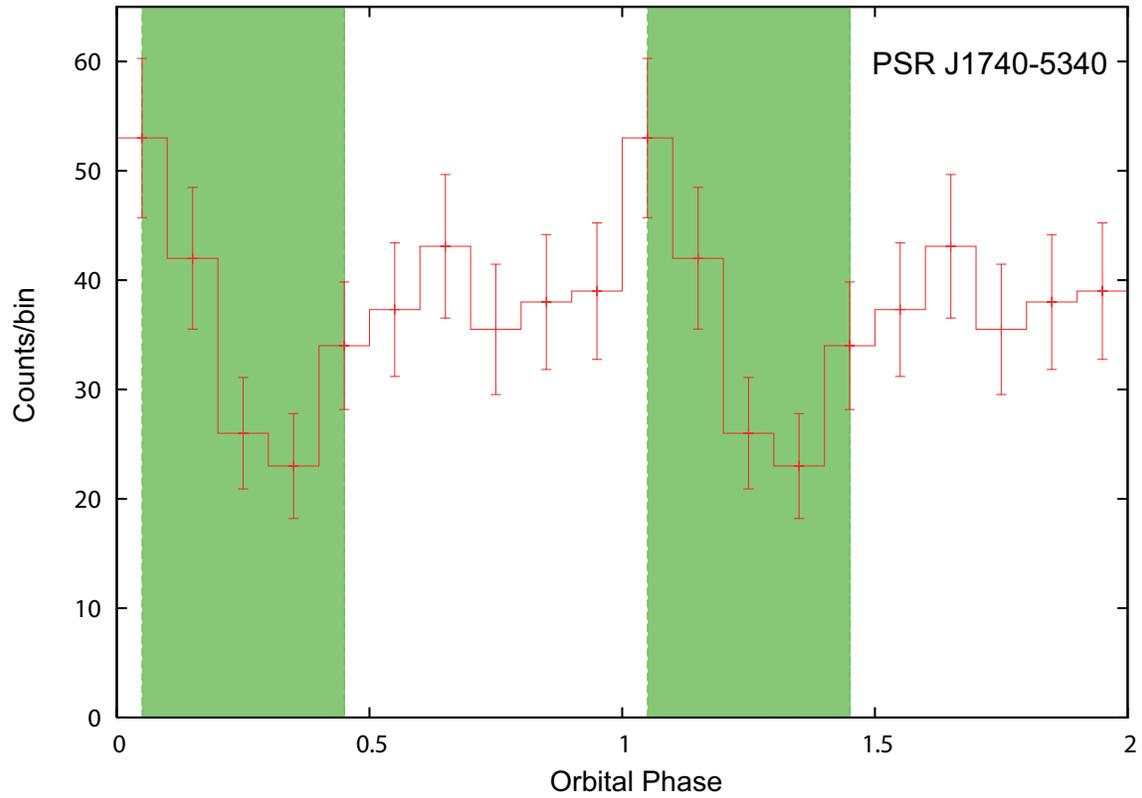,width=15cm,clip=}}
\caption[]{Lightcurve of PSR~J$1740-5340$. Two phase cycles are
 shown for clarity. The background noise level is found to be 
 at $\sim 0.35$ counts/bin.  $\phi = 0.0$ corresponds to the 
 ascending node of the pulsar orbit. Error bars indicate the
 $1\sigma$ uncertainty. The shaded  phase regions mark the 
 phases were the radio pulses eclipse.}
\label{figure15}
\end{figure}

The millisecond pulsar PSR~J1740-5340 in NGC~6397 is in a $\sim1.35$ day binary orbit with a 
massive late type companion. The pulsar's radio emission is seen to eclipse in the orbital 
phase interval $0.05-0.45$ \citep{DAmico2001}. Five datasets from NGC 6397 are available in
the Chandra archive (cf.~Table \ref{t:observations}), covering various phase 
ranges of the pulsar's binary orbit (see Huang \& Becker 2010 for a more detailed discussion). 
The first observation done in 2000 July 31 was aimed on the front-illuminated 
(FI) ACIS-I3 chip, while the other four observations were taken with the back-illuminated (BI) 
chip ACIS-S3. The pulsar's net counting rates obtained from this data are $(1.31\pm0.17)\times 
10^{-3}$ cts/s, $(1.74\pm 0.25)\times 10^{-3}$ cts/s, $(2.89\pm0.33)\times 10^{-3}$ cts/s, 
$(2.26\pm0.16)\times 10^{-3}$ cts/s and $(2.92\pm0.14)\times 10^{-3}$ cts/s, respectively, 
revealing a $\sim 3\sigma$ flux variability on time scales of  month to years. Figure \ref{figure15} 
shows the pulsar lightcurve as folded at the binary period. The significance of the flux modulation 
over the observed orbit was found to be between 88.5\% and 99,6\%, depending on the number of phase 
bins used to construct the lightcurve (cf.~Huang \& Becker 2010). 

\subsection{SPECTRAL ANALYSIS}

The ACIS-S CCDs provide spectral information for all cluster pulsars for which counterparts are
detected. The quality of spectral fits and the constraint on the parameters of the model-spectra, 
however, is a function of photon statistics and thus varies significantly among the detected 
counterparts. In addition, the column absorption towards the various globular clusters shows 
a large diversity (cf.~Table \ref{t:clusterParameter}), meaning that the sensitivity to detect 
soft X-rays from e.g.~thermal hot-spots on the neutron star surface may be higher for some 
clusters (e.g.~in 47 Tuc) than in others (e.g.~Terzan 5). 

The extraction of the source and background spectra as well as the computation of the 
corresponding response and effective area files were performed with the data reduction 
package CIAO~4.2 by using the calibration data in CALDB~4.2. To extract the spectra we 
used circular regions of $2''$ radius (corresponding to $\sim 95$\% encircled energy)
centered at the radio pulsar's timing position. Net counting rates for the pulsar counterparts 
(cf.~Table \ref{t:clusterParameter}) were obtained by subtracting an averaged background 
rate which we obtained from combining two different source-free regions located near to 
the pulsar. Power law and blackbody model spectra were then fitted to the extracted 
spectra by using XSPEC~12.3.1. All datasets of a pulsar counterpart were fitted 
simultaneously unless a temporal counting rate variability was found for e.g.~binary 
pulsars.

Most millisecond pulsar counterparts are detected with a relatively small photon 
statistics, requiring to fix the hydrogen column densities in order to better 
constrain the remaining model parameters. If so we deduced $N_{H}$ from 
the optical foreground reddening E(B$-$V) of the corresponding globular clusters 
(Harris 1996, updated 2003\footnote{http://physwww.mcmaster.ca/$\sim$harris/Databases.html}). 
Only for PSR~J1824-2452A in M28 and the two recent observations \#7460 and \#7461 of 
PSR~J1740-5340 in NGC~6397 the photon statistics was sufficient to include $N_{H}$ as 
an additional free parameter in the spectral fits. 

For all 47 Tuc pulsars we grouped the extracted spectra so as to have at least 5 counts 
per spectral bin. 47 Tuc F+S and G+I were treated as single sources since their spectra 
could not be separated. The spectral parameters for most of the 47 Tuc pulsars are
consistent with those reported by Bogdanov et al.~(2006), except for 47 Tuc L and R. 
The spectral fits of these two pulsars are not well described by a single blackbody 
model albeit different background regions were included or $N_{H}$ was let a free 
parameter in the fits. A composite model consisting of a blackbody plus power-law model
yields a better but still not optimal description of the observed spectra.

In order to investigate whether the binary pulsars PSR~J1740$-$5340 in NGC~6397 and 
J$1748-2021$B in NGC~6440 show spectral changes along with their variable counting rates 
observed at different orbital phase angles we analyzed each of their available 
datasets separately. As can be seen in Table~\ref{t:gc_psr_spec}, the photon indices obtained 
for a pulsar from each of the various datasets are all in agreement with each other 
(cf.~also Huang \& Becker 2010 for a more detailed discussion on PSR~J1740$-$5340 
in NGC~6397). The X-ray luminosity of PSR~J1748$-$2021B in NGC\,6440, though, changed by 
a factor of $\sim3$ for observations taken in 2000 July and 2003 June. 
PSR~J1748$-$2021B furthermore shares the non-thermal nature of its X-ray spectrum 
with PSR~J1953+1846A in M71 (cf.~Elsner et al.~2008) and PSR~J1701$-$3006C in M62. 
Their photon indices are in the range $\sim 1.4 - 1.9$. For the binary pulsars PSR~J1824-2452H 
in M28 and PSR~J1748-2446 in Terzan~5 the limited photon statistics did not support 
to distinguish between thermal or non-thermal emission models. A blackbody and a 
power-law model both describe their spectra with comparable goodness (cf.~Table \ref{t:gc_psr_spec}). 
The fitted blackbody temperatures, however, are at the level of $\gtrsim10^{7}~K$,
which is an order of magnitude higher than what is observed in other millisecond 
pulsar spectra (cf.~Table 6.7 in Becker 2009). A non-thermal interpretation of their 
emission therefore seems more likely to us.  Whether this non-thermal X-ray radiation 
is due to magnetospheric emission from within the pulsars' co-rotating light-cylinder or
whether it arrises because of an intra-binary shock formed by the relativistic pulsar
wind and the matter from the stellar companion is currently not clear.

For the brightest of all millisecond pulsars, PSR~J1824-2452A, five observations
are currently available in the Chandra data archive. Three of them were taken in 2002
and provided the basis for the results published on M28 by Becker et al.~(2003). Two 
additional and significantly longer observations of M28 were taken in 2008, increasing 
the total on source exposure to 237 ksec, i.e.~almost five times as much as was 
available in 2002. To analyze the spectrum of PSR~J1824-2452A we made use of all
five datasets.

\begin{figure}
 \centerline{\psfig{figure=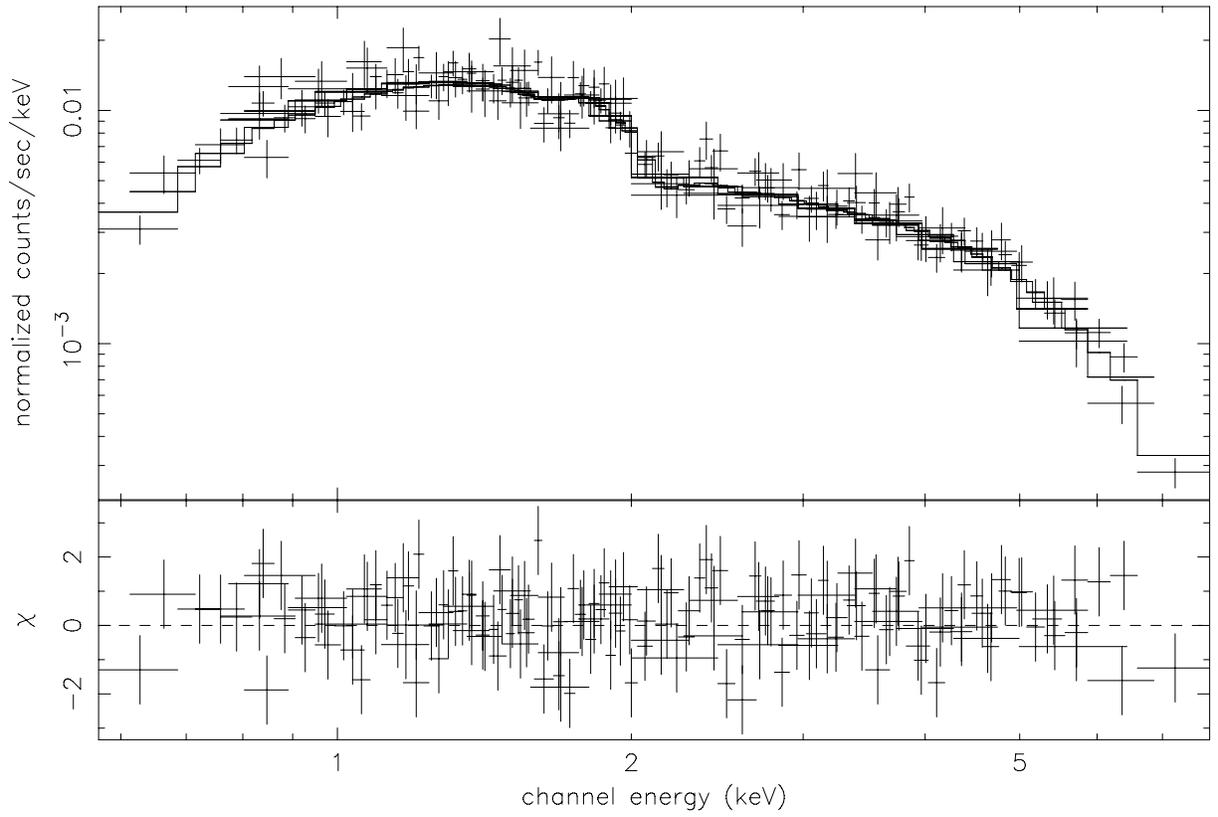,width=16cm,clip=}}
 \caption[]{Energy spectrum of the millisecond pulsar PSR J1824-2452A in M28
 obtained from the Chandra observations taken in 2002 and 2008 and fit to an 
 absorbed power-law model ({\it upper panel}) and contribution to the $\chi^2$ fit 
 statistic ({\it lower panel}).}
\label{figure16}
\end{figure}

The spectra based on the 2002 data were grouped so as to have at least 30 counts per 
spectral bin. For the 2008 data we used a grouping of 50 and 30 counts per spectral 
bin for the longer (\#9132) and shorter (\#9133) observations (cf.~Table \ref{t:observations}). 
It has been shown previously that the pulsar's X-radiation is dominated by non-thermal 
emission (Kawai \& Saito 1999, Becker et al.~2003, Mineo et al.~2004). Fitting a power-law 
spectral model to the data yields $N_{H}=(0.22\pm0.02)\times 10^{22}\,\mbox{cm}^{-2}$, 
a photon-index  $\alpha = 1.13^{+0.03}_{-0.04}$, and a normalization at 1 keV of 
$3.43^{+0.23}_{-0.15}\times10^{-5}$ photons cm$^{-2}$ s$^{-1}$ keV$^{-1}$ 
($\chi_\nu^2= 1.00$ for 168 dof). The column density is fully consistent with 
what is deduced from the reddening towards M28. The unabsorbed energy 
flux in the $0.3-8.0$ keV band is $f_x=3.7^{+0.4}_{-0.3} \times 10^{-13}\,\,
{\rm ergs\,\,s}^{-1}\,{\rm cm}^{-2}$, yielding an X-ray luminosity of 
$L_x=1.36^{+0.15}_{-0.11} \times 10^{33}\,{\rm ergs\,\, s}^{-1}$. If transformed to 
the $0.1-2.4\,$ keV ROSAT band this corresponds to $L_x=3.76^{+0.33}_{-0.22}\times 
10^{32}\,{\rm ergs\,\, s}^{-1}$, which is comparable with the luminosity inferred from the 
ROSAT data (Verbunt 2001). These luminosities imply a rotational energy $\dot{E}$
to X-ray energy conversion factor of $L_{x,0.3-8.0\,keV}/\dot{E}= 6.1 \times 10^{-4}$ 
and $L_{x,0.1-2.4\,keV}/ \dot{E}= 2.1 \times 10^{-4}$, respectively. The phase averaged 
photon index obtained in our spectral fits is identical to the one deduced for PSR~J1824-2452A 
from the observations at pulse maximum using joint ASCA and RXTE data (Kuiper et 
al.~2003). This is in agreement with the interpretation that its fraction of pulsed
X-ray photons is  $85\%-100\%$. Figure~\ref{figure16} shows the pulsar spectrum as 
fitted to an absorbed power-law model.

Unlike for many other millisecond pulsars (cf.~Table 6.7 of Becker 2009 for a summary), 
modeling the X-ray spectrum of PSR~J1824-2452A with a power-law does not require any 
additional blackbody component (e.g.~associated with thermal emission from heated polar 
caps) to get an acceptable spectral fit. All combinations of blackbody normalizations 
and temperatures that were fitted along the power-law model gave reduced $\chi^2$-values 
which didn't indicate a higher likelihood for such a model than the fits to a single 
power-law. The F-test statistic for adding the extra blackbody spectral component to 
the power-law model, thus, is very low.

Nevertheless, the high photon statistics provided by the archival Chandra data allows 
us to constrain the temperature of a presumed thermal polar cap. Defining the size of 
the polar cap as the foot points of the neutron star's dipolar magnetic field, 
the radius of the polar cap area is given by $\rho=\sqrt{2\pi R^3/c P}$ with $R$ 
being the neutron star radius, c the velocity of light and P the pulsar rotation 
period (see e.g.~Michel 1991). For PSR~J1824-2452A, with a rotation period of 3.05 ms 
this yields a polar cap radius of $\rho\sim 2.62$ km. 

\begin{figure}
\centerline{\psfig{figure=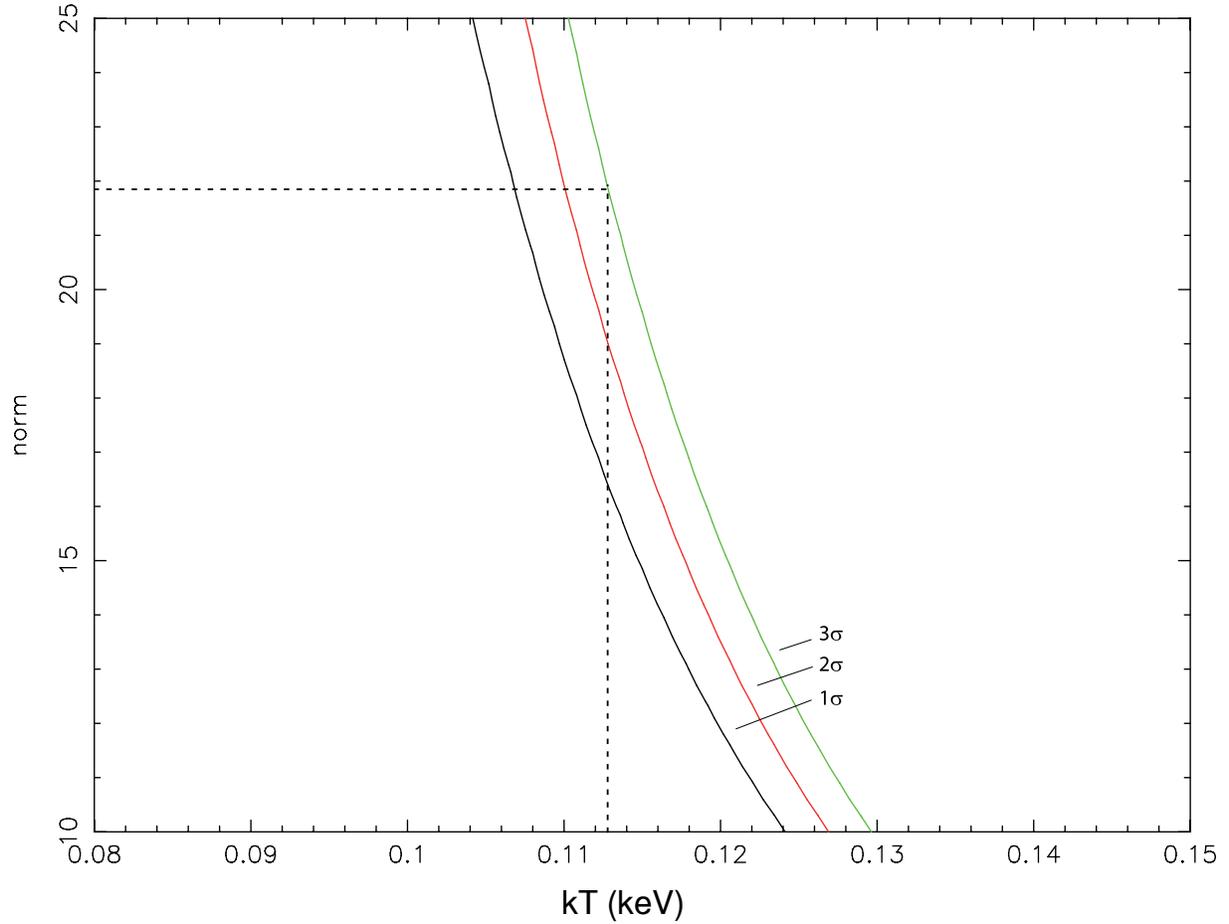,width=16cm,clip=}}
\caption[]{Portion of the confidence contours showing the blackbody
 normalization versus blackbody temperature for the composite model 
 (see text). The horizontal line at a normalization of 21.88 
 corresponds to a polar cap radius of $2.62$ km and a pulsar 
 distance of 5.6 kpc. The contours correspond to $\chi^2_{min}=167.5$ plus 
 2.3, 6.17 and 11.8 which are the $1\sigma$, $2\sigma$ and $3\sigma$ 
 confidence contours for 2 parameters of interest.}
\label{figure17}
\end{figure}

As a thermal spectral component of a heated polar cap contributes mostly below 
$\sim 1$ keV, the fitted column absorption is found to be a steep function of the 
blackbody emitting area (corresponding to the model normalization) and temperature. 
To determine a polar cap temperature upper limit which is in agreement with the 
fitted power-law model and column absorption we fixed the absorption of the composite 
model as well as the power-law photon index to the upper bound set by the $1\sigma$ 
confidence range deduced in the power-law fit. The power-law normalization was fixed
to the $1\sigma$ lower bound as this led to a higher temperature upper limit. We then 
computed the confidence ranges of the blackbody normalization and temperatures 
by leaving these parameters free. The resulting contours, computed for two 
parameters of interest, are shown in Figure \ref{figure17}. 

The blackbody normalization in XSPEC is proportional to $\rho^2_{km}/d^2_{10\,kpc}$ in 
which $\rho_{km}$ is the blackbody radius of the emitting area and $d_{10\,kpc}$ is the 
pulsar distance in units of 10 kpc. For a distance of 5.6 kpc towards M28 and a polar
cap radius of 2.62 km we thus obtain a normalization of 21.88. Assuming a contribution 
from one polar cap only we can set a $3\sigma$ temperature upper limit of 
$T_{pc}^\infty < 1.3 \times 10^6$ K. This upper limit is at the same level as the
temperatures fitted for the thermal components in the spectra of e.g.~the solitary 
millisecond pulsar PSR~J$2124-3358$ or of PSR~J$0437-471$5 (cf.~Table 6.7 in Becker 2009).
Converting the temperature upper limit into a flux upper limit yields 
$f_{bb, 0.3 - 8\,\,keV}\le 1.5 \times 10^{-14}\,\, {\rm ergs\,\,s}^{-1}\,{\rm cm}^{-2}$,
corresponding to $\le 4$\% of the non-thermal energy flux within $0.3-8$ keV.

The best-fit spectral models and parameters used to describe the spectra of the 
X-ray detected globular cluster millisecond pulsars are summarized in Table \ref{t:gc_psr_spec}.

\section{SUMMARY AND DISCUSSION}

We have systematically studied the X-ray emission properties of 31 globular cluster 
millisecond pulsar counterparts in order to evaluate their spectral properties 
and luminosities in a consistent way. Timing analysis has been performed for all 
sources according to the temporal resolution available in the archival Chandra data. 
Making use of unpublished data on M28, M4 and NGC 6752 allowed us to obtain further 
constraints for the millisecond pulsar counterparts located in these clusters. 
Counting rate and energy flux upper limits were computed for 36 globular cluster 
millisecond pulsars for which no X-ray counterparts could be detected. 

By spring 2010 emission from 98 rotation-powered pulsars has been detected in the soft 
X-ray band. 51 of these sources belong to the group of galactic field pulsars whereas 
the other 47 sources are recycled pulsars. 31 of them are located in nine globular 
clusters. Coherent X-ray pulses are detected from 10 millisecond pulsars of which four 
are in a globular cluster. These are PSR J0024-7204\,D,O,R in 47 Tuc for which we measured 
a pulsed fraction of $\sim 60\%$ and PSR J1824-2452A in M28 which has $\sim 85\%$ of its X-rays pulsed. 
Their X-ray pulse profiles are all characterized by two peaks, which in the case of PSR J1824-2452A 
are narrow (phase width  $\sim 10^\degree$ and $\sim 20^\degree$ (FWHM) for the first and second 
peak) and broader (typical phase width $\sim 100^\degree$ and $\sim 150^\degree$ (FWHM)) in the 
case of J0024-7204\,D,O,R. 

Comparing the X-ray lightcurves with the corresponding radio profiles shows that there 
is some correspondence between both. For PSR J1824-2452A the main X-ray pulse component appears 
to be almost phase aligned with the main radio pulse. No correspondence of the remaining two 
peaks seen at 800 MHz is found in the X-ray profile. The 47 Tuc pulsars J0024-7204\,D,O 
have two radio peaks observed at 1400 MHz (Freire et al., 2010, in prep.), albeit their second 
radio peak has a much lower intensity than the main peak. PSR J0024-7204D shows an asymmetry 
in the main radio peak as the X-ray peaks do. The profile of PSR J0024-7204\,R is characterized 
by a narrow X-ray peak followed by a broader pulse component covering the remaining rotation 
phase. The profile at 1400 MHz shows a striking gross similarity, consisting of a narrow main 
pulse and a broader pulse component which by itself has a sub-structure (Freire et al., 2010, 
in prep.). Clearly, a better photon statistics is required in order to establish this correspondence.

36 of the 67 millisecond pulsars which we have considered in this work are in binaries. 
X-ray flux variability either on time scales comparable with the pulsar's orbit period 
or even longer was found for only three of them; the binary pulsars PSR J0024-7204W in 
47 Tuc, PSR J1748-2021B in NGC 6440 and PSR J1740-5340 in NGC 6397. Some low-significant 
evidence for a flux variability along their binary motion was found for PSR J1824-2452H 
and J1824-2452G. 

Spectral information is available from all 31 globular cluster millisecond pulsar counterparts
(cf.~Table \ref{t:gc_psr_spec}). The complexity of the spectral models which could be tested 
and the accuracy of the fitted spectral parameters, however, are strongly inhomogeneous among 
all of them. This is not only because of a diversity of the column absorption towards the 
various clusters but also because of varying photon statistics, which for fainter sources 
often do not even allow to distinguish between simple blackbody and power-law spectral models. 

PSR J1824-2452A is the brightest among all millisecond pulsars and is a non-thermal Crab-like 
X-ray emitter. Using the 2002 Chandra data, Becker et al.~(2003) found in its X-ray spectrum 
some evidence for a possible line feature at $\sim 3.3$\,keV (cf.~Figure~2 in Becker et al.~2003). 
The new data taken in 2008 supersede the previous data in photon statistics and sensitivity and 
do not confirm the presence of this feature. A $3\sigma$ temperature upper limit of $1.3\times 
10^6$ K is deduced for a presumed thermal polar cap contribution. This is at the same level 
as observed in spectral fits of other millisecond pulsars. That non-thermal radiation is the 
dominating emission component in PSR~J1824-2452A thus does not exclude the existence of a 
thermal polar cap of similar properties than observed in e.g.~PSR J0437-4715 or PSR J2124-3358 
(cf.~Table 6.7 in Becker 2009). The 47 Tuc pulsars all fit relatively well with a thermal 
spectrum or a composite spectral model consisting of a blackbody and power-law. The latter 
model especially fits the spectra of e.g.~PSR J0024-7204\,O,R for which X-ray pulses are detected. 

Grindlay et al.~(2001) suggested that the X-ray efficiency for the 47~Tuc millisecond pulsars 
is $L_{\rm X} \propto \dot{E}^{0.48\pm0.15}$. Such dependence is obviously less steep than the 
linear relation $L_{\rm X} \sim 10^{-3} \dot{E}$ found for the isotropic X-ray luminosity of
rotation-powered pulsars by Becker \& Tr\"umper (1997). Bogdanov et al.~(2006) found a relation 
of $L_{\rm X}\propto\dot{E}^{0.2\pm1.1}$ for the 47\,Tuc pulsars using more significant Chandra 
data taken in September/October 2002. We have reexamined the isotropic X-ray conversion efficiency 
of the globular cluster pulsar population in adding the results reported in this paper into the 
data pool. To compute the spin-down power $\dot{E}$ of the globular cluster pulsars we used 
the estimated intrinsic spin-down rate $\dot{P}$ reported in the literature. Pulsars which 
turned out to be variable were excluded from the correlation. As both, emission from heated 
polar caps and non-thermal emission finally appears to be powered by the rotation of the star 
the X-ray luminosities from both spectral components were added and correlated with $\dot{E}$. 
Adopting the absorption-corrected isotropic X-ray luminosity listed in Table \ref{t:gc_psr_spec} 
and its statistical 1-$\sigma$ error we fitted $\log L_{X}=(14.49 \pm 4.47) + (0.48 \pm 0.11) 
\log \dot{E}$ with a correlation coefficient 0.68. 

\begin{figure}
\centerline{\psfig{figure=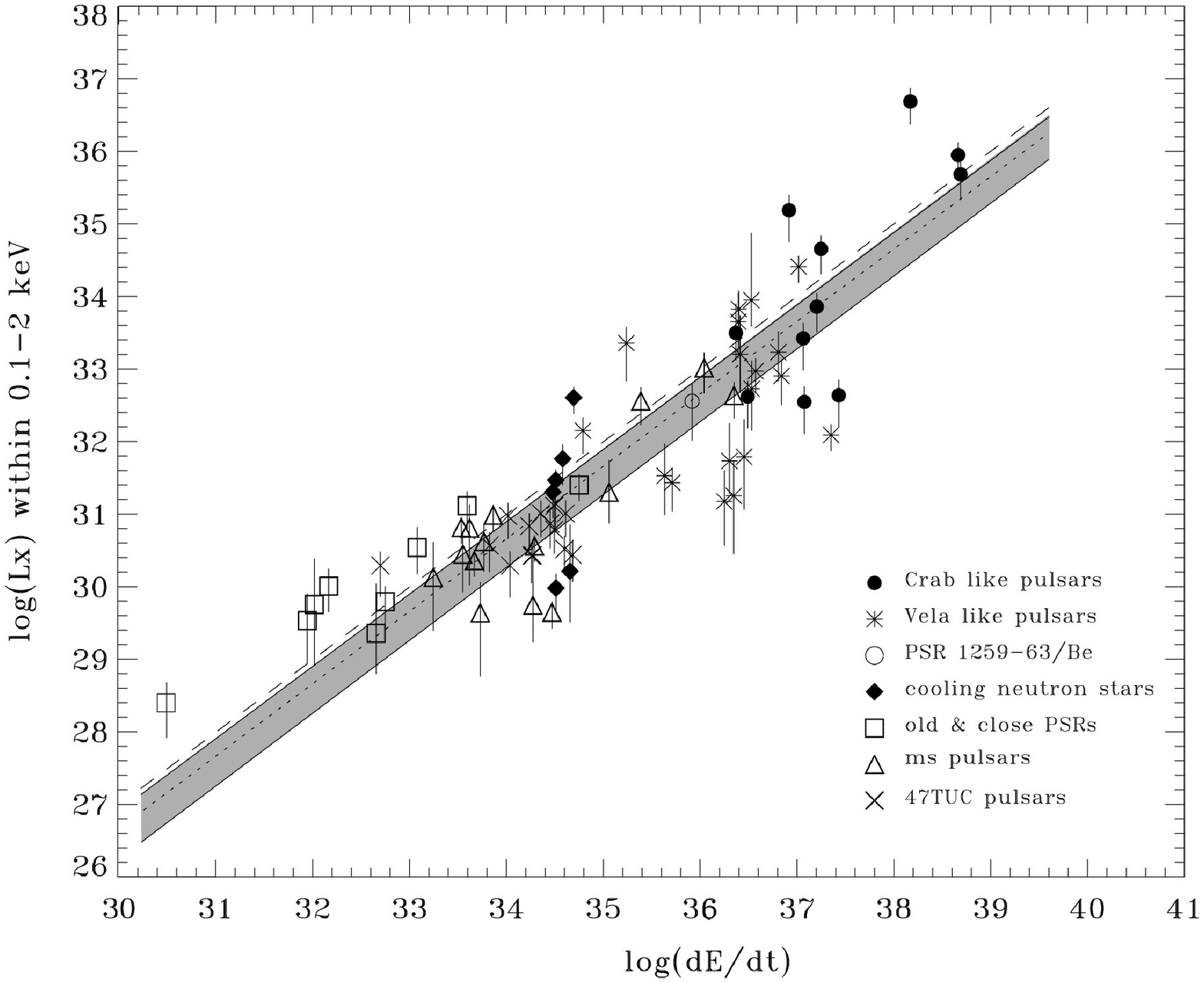,width=16cm,clip=}}
\caption[]{Isotropic X-ray luminosity vs.~spin-down energy of X-ray detected 
rotation-powered pulsars. (Status: spring 2010)}
\label{figure18}
\end{figure}

Correlating the spin-down energy and isotropic X-ray luminosity of $\sim 80$ rotation-powered
pulsars for which spectral information was obtained in Chandra and/or XMM-Newton observations 
Becker (2009) fitted  \[L_X (0.1-2\,\mbox{keV})= 10^{-3.24^{+0.26}_{-0.66}}\, \dot{E}^{0.997^{+0.008}_{-0.001}}\] 
in which the errors in $L_x$ have been fully taken into account and were used to weight the 
data points.  Although the data includes three times as much pulsars than were available
with ROSAT the correlation is still in good agreement with $L_{\rm X}(0.1-2.4\,\mbox{keV})
\sim 10^{-3}\, \dot{E}$ (cf.~Figure \ref{figure18}). With the larger database it now becomes 
evident that this relation represents an averaged approximation to the X-ray efficiency rather
than a fixed correlation. This was already suggested by Becker \& Tr\"umper (1997) 
and may be due to the fact that ROSAT with its limited sensitivity was able to detect essentially 
only the brightest pulsars.  With the higher sensitivity of XMM-Newton and Chandra more faint pulsars 
could be detected in which e.g.~the orientation of their magnetic/rotational axes to the 
observers line of sight might not have been optimal. As no beaming correction can be applied 
to the observed luminosities the X-ray efficiency of those pulsars appears to be smaller,
even though their spin-down energy might be comparable to more efficient emitters. 

\begin{figure}
\centerline{\psfig{figure=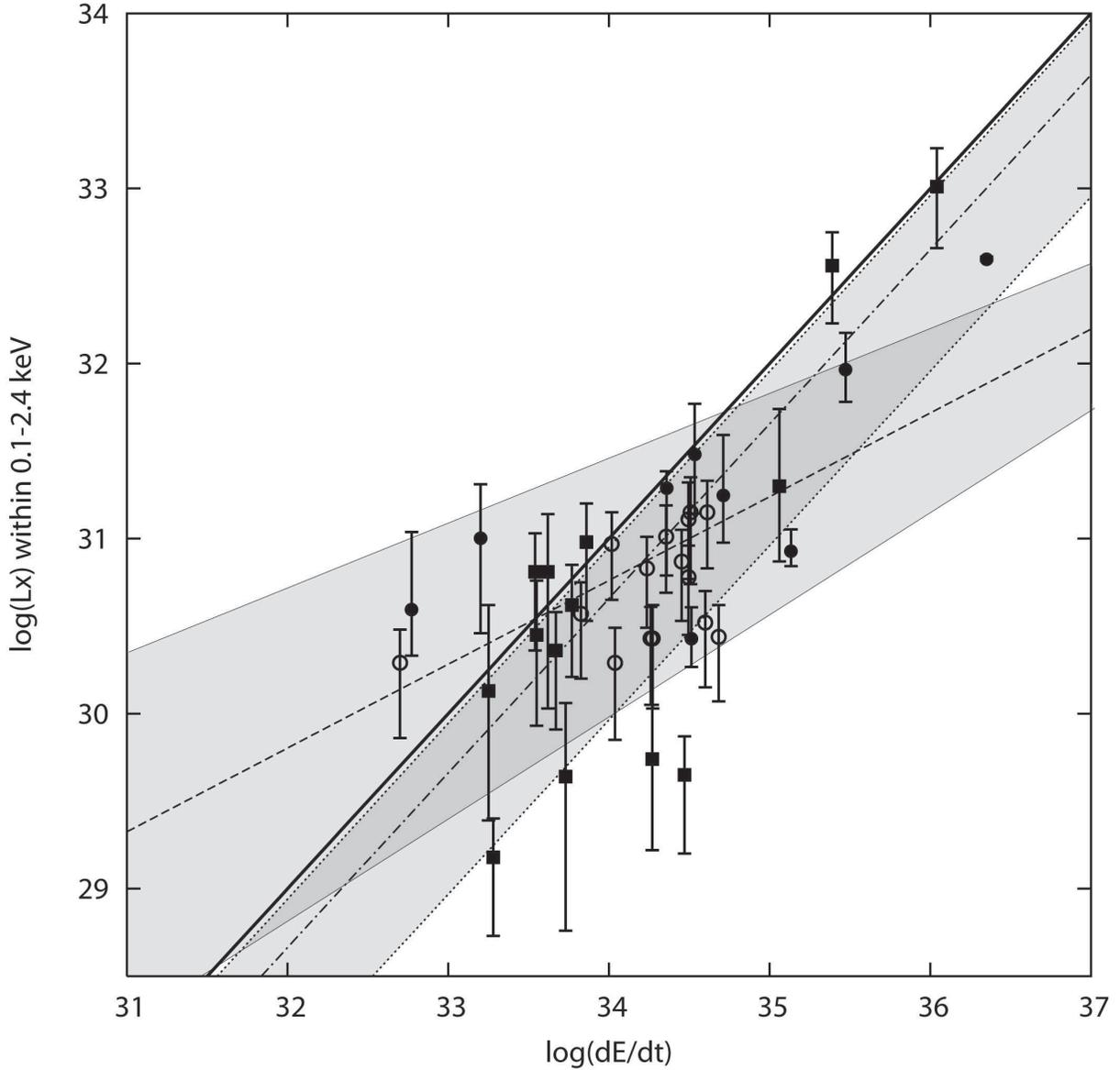,width=16cm,clip=}}
\caption[]{Isotropic X-ray luminosity vs.~spin-down energy of X-ray detected millisecond pulsars.
Galactic field millisecond pulsars are represented by {\it filled squares}, 47\,Tuc pulsars
by {\it open circles} and other globular cluster pulsars by {\it filled circles}.
The solid line represents $L_\mathrm{X, 0.1-2.4\,keV} = 10^{-3} \dot{E}$ (Becker \& Tr\"umper 1997) 
while the dot-dashed and dotted lines with its gray-shaded $1\sigma$ uncertainty range 
is from a fit to all X-ray detected millisecond pulsars (Becker 2009; cf.~Figure \ref{figure15}). 
The dashed line and its gray-shaded $1\sigma$ uncertainty range is from correlating the spin-down 
energy and the X-ray luminosities of the globular cluster millisecond pulsars only.}
\label{figure19}
\end{figure}

As far as the X-ray efficiency from the globular cluster millisecond pulsars is concerned it can 
be seen from Figure \ref{figure19} that their X-ray luminosities are well within the scatter of 
other data points at this spin-down energy level. Within the uncertainties of the deduced X-ray 
luminosities it is therefore not justified to conclude that these pulsars have an X-ray efficiency 
which is different from the one observed for e.g.~field millisecond pulsars. Millisecond 
pulsars in the galactic plane and in globular clusters thus appear to show no distinct differences 
in their X-ray emission properties.

\acknowledgments
\noindent{\bf Acknowledgments}\linebreak
We thank J.~Tr\"umper for his comments and P.~Freire for making the radio
profiles of PSR J0024--7204D, J0024--7204O and J0024--7204R available 
to us prior publication. H.H. Huang acknowledges support by the International 
Max-Planck Research School on Astrophysics, IMPRS. We acknowledge the use of
the Chandra data archive.

\clearpage

\clearpage

\clearpage

\clearpage

\clearpage

\clearpage

\clearpage

\clearpage

\clearpage

\clearpage

\clearpage

\clearpage

\clearpage

\clearpage

\clearpage

\clearpage

\clearpage

\clearpage

\clearpage

\clearpage

\begin{deluxetable}{lccccccc}                                            
\tabletypesize{\scriptsize}
\tablewidth{0pc}
\tablecaption{Chandra observations of globular clusters which are known to host millisecond pulsars. 
The table is ordered top-down according to the number of radio millisecond pulsars known in 
each cluster. (Status: spring 2010). \label{t:observations} \label{table1}}
\tablehead{                                                                
 Cluster            &Obs-ID&  Detector  & Start Date  &       Exposure                              \\
    {}              &  {}  &    {}      &    {}       &       (s)$\!\!\!\!$                         } 
\startdata                                                                                                                          
  Terzan 5          & 654  &   ACIS-I   & 2000-07-29  &       5510$^1\!\!\!\!\!\!\!\!\!\!$          \\                      
    {}              & 655  &   ACIS-I   & 2000-07-24  &       42161$^1\!\!\!\!\!\!\!\!$             \\                                   
    {}              & 3798 &   ACIS-S   & 2003-07-13  &       39344$\!\!\!\!$                       \\ \hline\\[-2ex]                    
47~Tucanea (NGC~104)&   78 &   ACIS-I   & 2000-03-16  &        4050$\!\!\!\!$ $\!\!\!\!\!\!\!$      \\                   
     {}             &  953 &   ACIS-I   & 2000-03-16  &       32080$\!\!\!\!$                       \\                                   
     {}             &  954 &   ACIS-I   & 2000-03-16  &         920$\!\!\!\!\,\!\!\!\!\!\!\!\!\!\!$ \\             
     {}             &  955 &   ACIS-I   & 2000-03-16  &       32080$\!\!\!\!$                       \\                                   
     {}             &  956 &   ACIS-I   & 2000-03-17  &        4910$\!\!\!\!\!\!\!\!\!$             \\                        
     {}             & 2735 &   ACIS-S   & 2002-09-29  &       66100$\!\!\!\!$                       \\                                   
     {}             & 2736 &   ACIS-S   & 2002-09-30  &       66110$\!\!\!\!$                       \\                                   
     {}             & 2737 &   ACIS-S   & 2002-10-02  &       66110$\!\!\!\!$                       \\                                   
     {}             & 2738 &   ACIS-S   & 2002-10-11  &       69860$\!\!\!\!$                       \\                                   
     {}             & 3384 &   ACIS-S   & 2002-09-30  &        5580$\!\!\!\!\!\!\!\!\!$             \\                                   
     {}             & 3385 &   ACIS-S   & 2002-10-01  &        5580$\!\!\!\!\!\!\!\!\!$             \\                                   
     {}             & 3386 &   ACIS-S   & 2002-10-03  &        5830$\!\!\!\!\!\!\!\!\!$             \\                                   
     {}             & 3387 &   ACIS-S   & 2002-10-11  &        6030$\!\!\!\!\!\!\!\!\!$             \\                                   
     {}             & 5542 &   HRC-S    & 2005-12-19  &       50160$\!\!\!\!$                       \\                                   
     {}             & 5543 &   HRC-S    & 2005-12-20  &       51390$\!\!\!\!$                       \\                                   
     {}             & 5544 &   HRC-S    & 2005-12-21  &       50140$\!\!\!\!$                       \\                                   
     {}             & 5545 &   HRC-S    & 2005-12-23  &       51870$\!\!\!\!$                       \\                                   
     {}             & 5546 &   HRC-S    & 2005-12-27  &       50150$\!\!\!\!$                       \\                                   
     {}             & 6230 &   HRC-S    & 2005-12-28  &       49400$\!\!\!\!$                       \\                                   
     {}             & 6231 &   HRC-S    & 2005-12-29  &       47150$\!\!\!\!$                       \\                                   
     {}             & 6232 &   HRC-S    & 2005-12-31  &       44360$\!\!\!\!$                       \\                                   
     {}             & 6233 &   HRC-S    & 2006-01-02  &       97930$\!\!\!\!$                       \\                                   
     {}             & 6235 &   HRC-S    & 2006-01-04  &       50130$\!\!\!\!$                       \\                                   
     {}             & 6236 &   HRC-S    & 2006-01-05  &       51920$\!\!\!\!$                       \\                                   
     {}             & 6237 &   HRC-S    & 2005-12-24  &       50170$\!\!\!\!$                       \\                                   
     {}             & 6238 &   HRC-S    & 2005-12-25  &       48400$\!\!\!\!$                       \\                                   
     {}             & 6239 &   HRC-S    & 2006-01-06  &       50160$\!\!\!\!$                       \\                                   
     {}             & 6240 &   HRC-S    & 2006-01-08  &       49290$\!\!\!\!$                       \\ \hline\\[-2ex]                    
M28 (NGC~6626)      & 2683 &   ACIS-S   & 2002-09-09  &       14110$\!\!\!\!$                       \\                               
   {}               & 2684 &   ACIS-S   & 2002-07-04  &       12746$\!\!\!\!$                       \\                                   
   {}               & 2685 &   ACIS-S   & 2002-08-04  &       13511$\!\!\!\!$                       \\                                   
   {}               & 9132 &   ACIS-S   & 2008-08-07  &       142260                                \\                                   
   {}               & 9133 &   ACIS-S   & 2008-08-10  &       54456$\!\!\!\!$                       \\                     
   {}               & 2797 &   HRC-S    & 2002-11-08  &       49370$\!\!\!\!$                       \\
   {}               & 6769 &   HRC-S    & 2006-05-27  &       41070$\!\!\!\!$                       \\ \hline\\[-2ex]
M15 (NGC~7078)     & 1903  &    HRC-I   & 2001-07-13  &        9096$\!\!\!\!\!\!\!\!$               \\                       
   {}              & 2412  &    HRC-I   & 2001-08-03  &        8821$\!\!\!\!\!\!\!\!$               \\                            
   {}              & 2413  &    HRC-I   & 2001-08-22  &       10790$\!\!\!\!$                       \\                            
   {}              & 9584  &    HRC-I   & 2007-09-05  &       21445$\!\!\!\!$                       \\ \hline\\[-2ex]             
 NGC~6440          &  947  &   ACIS-S   & 2000-07-04  &       23279$\!\!\!\!$                       \\                              
   {}              & 3799  &   ACIS-S   & 2003-06-27  &       24046$\!\!\!\!$                       \\ \hline\\[-2ex]                   
M62 (NGC~6266)     & 2677  &   ACIS-S   & 2002-05-12  &       62266$\!\!\!\!$                       \\ \hline\\[-2ex]              
  NGC~6752         &  948  &   ACIS-S   & 2000-05-12  &       29468$\!\!\!\!$                       \\
    {}             & 6612  &   ACIS-S   & 2006-02-10  &       37967$\!\!\!\!$                       \\ \hline\\[-2ex]
 M3 (NGC 5272)     & 4542  &   ACIS-S   & 2003-11-11  &        9932$\!\!\!\!\!\!\!\!$               \\ 
    {}             & 4543  &   ACIS-S   & 2004-05-09  &       10152$\!\!\!\!$                       \\       
    {}             & 4544  &   ACIS-S   & 2005-01-10  &        9441$\!\!\!\!\!\!\!\!$               \\ \hline\\[-2ex]
 M5 (NGC~5904)     & 2676  &   ACIS-S   & 2002-09-24  &       44656$\!\!\!\!$                       \\\hline\\[-2ex]    
M13 (NGC~6205)     & 5436  &   ACIS-S   & 2006-03-11  &       26799$\!\!\!\!$                       \\                  
    {}             & 7290  &   ACIS-S   & 2006-03-09  &       27895$\!\!\!\!$                       \\ \hline\\[-2ex]       
 NGC~6441          &   721 &    HRC-I   & 2000-05-23  &        2290$\!\!\!\!\!\!\!\!$               \\\hline\\[-2ex]
 M22 (NGC~6656)    &  5437 &   ACIS-S   & 2005-05-24  &       16020$\!\!\!\!$                       \\\hline\\[-2ex]        
 M30 (NGC~7099)    &  2679 &   ACIS-S   & 2001-11-19  &       49435$\!\!\!\!$                       \\\hline\\[-2ex]        
 NGC~6544          &  5435 &   ACIS-S   & 2005-07-20  &       16278$\!\!\!\!$                      \\\hline\\[-2ex]        
 M4 (NGC~6121)     &   946 &   ACIS-S   & 2000-06-30  &       25816$\!\!\!\!$                      \\                      
     {}            &  7446 &   ACIS-S   & 2007-07-06  &       46040$\!\!\!\!$                      \\                          
     {}            &  7447 &   ACIS-S   & 2007-09-18  &       48540$\!\!\!\!$                      \\\hline\\[-2ex]            
 M53 (NGC~5024)    &  6560 &   ACIS-S   & 2006-11-13  &       24565$\!\!\!\!$                      \\\hline\\[-2ex]       
 M71 (NGC~6838)    &  5434 &   ACIS-S   & 2004-12-20  &       52446$\!\!\!\!$                      \\\hline\\[-2ex]        
 NGC~6397          &    79 &   ACIS-I   & 2000-07-31  &       48343$\!\!\!\!$                      \\                     
    {}             &  2668 &   ACIS-S   & 2002-05-13  &       28100$\!\!\!\!$                      \\                         
    {}             &  2669 &   ACIS-S   & 2002-06-30  &       26660$\!\!\!\!$                       \\                         
    {}             &  7460 &   ACIS-S   & 2007-07-16  &       147710                               \\                         
    {}             &  7461 &   ACIS-S   & 2007-06-22  &       88898$\!\!\!\!$                      \\\hline                   
\enddata                                                                                                                     
\tablecomments{\tiny $^1$Data unusable because of a bright flaring source in the field of view.}                             
\end{deluxetable}                                                                                                   

\clearpage

\begin{deluxetable}{lcccc}
\tabletypesize{\scriptsize}
\tablewidth{0pc}
\tablecaption{Distance, half-mass radius and column density of globular clusters which are
known to host millisecond pulsars. The table is ordered top-down according to the number of radio
millisecond pulsars known in each cluster. (Status: spring 2010).\label{t:clusterParameter} \label{table2}}
\tablehead{
 Cluster            & Distance & $r_{hm}$ &  $r_{core}$    & N$_{H}$  \\
    {}              & (kpc)    & (arcmin) &    (arcmin)    & ($10^{21}\mbox{\,cm}^{-2}$) }
\startdata
  Terzan 5          & 10.3     & 0.83     &   0.18   & 12.0   \\
47~Tucanea (NGC~104)& 4.5      & 2.79     &   0.44   & 0.1    \\
M28 (NGC~6626)      & 5.6      & 1.56     &   0.24   & 2.2    \\
M15 (NGC~7078)      & 10.3     & 1.06     &   0.07   & 0.6    \\
 NGC~6440           & 8.4      & 0.58     &   0.13   & 5.9    \\
M62 (NGC~6266)      & 6.9      & 1.23     &   0.18   & 2.6    \\
  NGC~6752          & 4.0      & 2.34     &   0.17   & 0.2    \\
 M3 (NGC 5272)      & 10.4     & 1.12     &   0.55   & 0.06   \\
 M5 (NGC~5904)      & 7.5      & 2.11     &   0.42   & 0.2    \\
M13 (NGC~6205)      & 7.7      & 1.49     &   0.78   & 0.1    \\
 NGC~6441           & 11.7     & 0.64     &   0.11   & 2.6    \\
 M22 (NGC~6656)     & 3.2      & 1.10     &   0.77   & 1.9    \\
 M30 (NGC~7099)     & 8.0      & 1.15     &   0.06   & 0.2    \\
 NGC~6544           & 2.7      & 1.77     &   0.05   & 4.1    \\
 M4 (NGC~6121)      & 2.2      & 3.65     &   0.83   & 2.0    \\
 M53 (NGC~5024)     & 17.8     & 1.11     &   0.36   & 0.1    \\
 M71 (NGC~6838)     & 4.0      & 1.65     &   0.63   & 1.4    \\
 NGC~6397           & 2.3      & 2.33     &   0.05   & 1.0    \\\hline
\enddata
\tablecomments{\tiny Globular cluster parameters from Harris (1996, 
update 2003). N$_{H}$ from the optical reddening E(B-V) of the
corresponding globular clusters.}
\end{deluxetable}

\clearpage

\begin{deluxetable}{l c c c c c}
\tabletypesize{\scriptsize}
\tablewidth{0pc}
\tablecaption{Period, spin-down energy and magnetic field strength of X-ray detected millisecond pulsars in
globular clusters. (Status: spring 2010). \label{t:ms-detected-basicproperties}\label{table3}}
\tablehead{
  Cluster                   &     Pulsar    &$P$~$^{a}$  &   $\dot{E}$           & $B_\perp$    & Binary$^{b}$\\
     {}                     &       {}      &$10^{-3}$ s &  ergs~s$^{-1}  $      & $10^{9}$ G   &   {}}
\startdata
47~Tucanea$^{c}$ (NGC~104)  &  J0024-7204C  &  5.7568    &  $5\times10^{32}$     &      -       &   N   \\        
       {}                   &  J0024-7204D  &  5.3576    &  $6.7\times10^{33}$   &      -       &   N   \\
       {}                   &  J0024-7204E  &  3.5363    &  $3.1\times10^{34}$   &     0.60     &   Y   \\
       {}                   &  J0024-7204F  &  2.6236    &  $4.1\times10^{34}$   &     0.42     &   N   \\
       {}                   &  J0024-7204G  &  4.0404    &  $<1.7\times10^{34}$  &      -       &   N   \\
       {}                   &  J0024-7204H  &  3.2103    &          -            &      -       &   Y   \\
       {}                   &  J0024-7204I  &  3.4850    &  $<7.1\times10^{34}$  &      -       &   Y   \\
       {}                   &  J0024-7204J  &  2.1006    &  $3.2\times10^{34}$   &      -       &   Y   \\
       {}                   &  J0024-7204L  &  4.3462    &  $1.0\times10^{34}$   &      -       &   N   \\
       {}                   &  J0024-7204M  &  3.6766    &          -            &      -       &   N   \\
       {}                   &  J0024-7204N  &  3.0540    &  $1.9\times10^{34}$   &      -       &   N   \\
       {}                   &  J0024-7204O  &  2.6433    &  $3.1\times10^{34}$   &     0.29     &   Y   \\
       {}                   &  J0024-7204Q  &  4.0332    &  $1.8\times10^{34}$   &     0.38     &   Y   \\
       {}                   &  J0024-7204R  &  3.4805    &  $2.8\times10^{34}$   &      -       &   Y   \\
       {}                   &  J0024-7204S  &  2.8304    &  $2.3\times10^{34}$   &      -       &   Y   \\
       {}                   &  J0024-7204T  &  7.5885    &  $1.1\times10^{34}$   &     1.51     &   Y   \\
       {}                   &  J0024-7204U  &  4.3428    &  $4.0\times10^{34}$   &     0.65     &   Y   \\
       {}                   &  J0024-7204W  &  2.3523    &         -             &      -       &   Y   \\
       {}                   &  J0024-7204Y  &  2.1967    &  $4.8\times10^{34}$   &      -       &   Y   \\\\[-1ex]

 M28 (NGC 6626)             &  J1824-2452A  &  3.0543    &  $2.2\times10^{36}$   &     2.25     &   N   \\
       {}                   &  J1824-2452G  &  5.9091    &  $3.4\times10^{34}$   &     1.07     &   Y   \\
       {}                   &  J1824-2452H  &  4.6294    &  $3.3\times10^{34}$   &     0.64     &   Y   \\\\[-1ex]

   NGC 6440                 & J1748-2021B   & 16.7601    &  $2.8\times10^{33}$   &     2.45     &   Y   \\\\[-1ex]

  M62 (NGC 6266)            & J1701-3006B   & 3.5939     &  $3.0\times10^{35}$   &     1.17     &   Y   \\
       {}                   & J1701-3006C   & 3.8064     &  $2.3\times10^{34}$   &     0.36     &   Y   \\\\[-1ex]

    NGC 6752                & J1911-6000C   & 5.2773     & $5.9\times10^{32}$    &     0.11     &   N   \\
       {}                   & J1910-5959D   & 9.0353     & $5.2\times10^{34}$    &     3.08     &   N   \\\\[-1ex]

   M4 (NGC 6121)            & B1620-26      & 11.0758    & $1.6\times10^{33}$    &     0.81     &   Y$^{d}$ \\\\[-1ex]

   M71 (NGC 6838)           & J1953+1846A   & 4.888      &         -             &       -      &   Y   \\\\[-1ex]

    Terzan 5                & J1748-2446    &   -        &         -             &       -      &   -   \\\\[-1ex]

    NGC 6397                & J1740-5340    &  3.6503    & $3.3\times10^{34}$    &      0.82    &   Y   \\\hline
\enddata
\tablecomments{\tiny
a. From http://www.naic.edu/$\sim$pfreire/GCpsr.html. b. Indicates whether the pulsar is in a binary system.\\
c. $\dot{E}$ from Table 4 of Bogdanov et al.~(2006). d. Long-orbit triple system with a white dwarf and a planet.}
\end{deluxetable}

\clearpage

\begin{deluxetable}{l c c c c c}
\tabletypesize{\scriptsize}
\tablewidth{0pc}
\tablecaption{Period, spin-down energy and magnetic field strength of X-ray non-detected millisecond pulsars in globular clusters.(Status: spring 2010).
\label{t:ms-not-detected-basicproperties}\label{table4}}
\tablehead{
 Cluster                   &     Pulsar    &$P$~$^{a}$  &   $\dot{E}$           & $B_\perp$    & Binary$^{b}$\\
    {}                     &       {}      &$10^{-3}$ s &  ergs~s$^{-1}  $      & $10^{9}$ G   &   {}}
\startdata

NGC\,6440        & J1748-2021A    &   288.603       &  $6.6\times10^{32}$     &  344      &   N      \\
  {}             & J1748-2021C    &    6.2269       &          -              &   -       &   N      \\
  {}             & J1748-2021D    &   13.4958       &  $9.4\times10^{33}$     &  2.85     &   Y      \\
  {}             & J1748-2021E    &   16.2640       &  $2.9\times10^{33}$     &  2.28     &   N      \\
  {}             & J1748-2021F    &    3.7936       &          -              &   -       &   Y      \\\\[-1ex]

NGC\,6441        & J1750-3703A    &   111.608       &  $1.6\times10^{32}$     &  25.4     &   Y      \\
  {}             & J1750-3703B    &    6.0745       &  $3.4\times10^{33}$     &  3.45     &   Y      \\
  {}             & J1750-3703C    &   26.5687       &          -              &   -       &   N      \\
  {}             & J1750-3703D    &    5.1399       &  $1.4\times10^{35}$     &  16.1     &   N      \\\\[-1ex]

NGC\,6544        & J1807-2459A    &    3.0595       &          -              &   -       &   Y      \\\\[-1ex]

NGC\,6752        & J1910-5959B    &    8.3578       &          -              &   -       &   N      \\
  {}             & J1910-5959E    &    4.5718       &          -              &   -       &   N      \\\\[-1ex]

M13 (NGC\,6205)  & J1641+3627A    &   10.3775       &          -              &   -       &   N      \\\\[-1ex]

M15 (NGC\,7078)  & B2127+11A      &   110.665       &          -              &   -       &   N      \\
  {}             & B2127+11B      &   56.1330       &  $2.1\times10^{33}$     &  23.4     &   N      \\
  {}             & B2127+11C      &   30.5293       &  $6.9\times10^{33}$     &  12.5     &   Y      \\
  {}             & B2127+11D      &    4.8028       &          -              &   -       &   N      \\
  {}             & B2127+11E      &    4.6514       &  $7.0\times10^{34}$     &  0.92     &   N      \\
  {}             & B2127+11F      &    4.0270       &  $1.9\times10^{34}$     &  0.36     &   N      \\
  {}             & B2127+11G      &   37.6602       &  $1.5\times10^{33}$     &  8.78     &   N      \\
  {}             & B2127+11H      &    6.7434       &  $3.1\times10^{33}$     &  0.41     &   N      \\\\[-1ex]

M28 (NGC\,6626)  & J1824-2452B    &    6.5466       &          -              &   $<0.4$  &   N      \\
  {}             & J1824-2452C    &    4.1583       &  $9.3\times10^{34}$     &   $<1.2$  &   Y      \\
  {}             & J1824-2452D    &   79.8354       &  $7.6\times10^{34}$     &  $\sim 91.0$& Y      \\
  {}             & J1824-2452E    &    5.4191       &          -              &   $<0.8$  &   N      \\
  {}             & J1824-2452F    &    2.4511       &  $2.5\times10^{34}$     &  $<0.5$   &   N      \\
  {}             & J1824-2452I    &    3.9318       &          -              &       -   &   Y      \\
  {}             & J1824-2452J    &    4.0397       &          -              &   $<0.6$  &   Y      \\\\[-1ex]

M30 (NGC\,7099)  & J2140-2310A    &   11.0193       &          -              &     -     &   Y      \\\\[-1ex]

M3 (NGC\,5272)   & J1342+2822B    &    2.389        &   $5.4\times10^{34}$    &  0.21     &   Y      \\
  {}             & J1342+2822D    &    5.443        &          -              &     -     &   Y      \\\\[-1ex]

M5 (NGC\,5904)   & B1516+02A      &    5.5536       &   $9.5\times10^{33}$    &  0.48     &   N      \\
  {}             & B1516+02B      &    7.9469       &          -              &     -     &   Y      \\\\[-1ex]

M62 (NGC\,6266)  & J1701-3006A    &    5.2416       &          -              &     -     &   Y      \\\\[-1ex]

Terzan 5         & J1748-2446A    &   11.5632       &          -              &     -     &   Y      \\
  {}             & J1748-2446C    &    8.4361       &          -              &     -     &   N      \\\hline
\enddata

\tablecomments{\tiny
a. From http://www.naic.edu/$\sim$pfreire/GCpsr.html. b. Indicates whether the pulsar is in a binary system.}
\end{deluxetable}

\pagebreak

\begin{deluxetable}{l l c c c c c c}
\tabletypesize{\scriptsize}           
\tablewidth{0pc}                      
\tablecaption{X-ray counterparts of globular cluster millisecond pulsars and their Chandra ACIS-S counting rates. (Status: spring 2010).
\label{t:xraydetections}\label{table5}}     
\tablehead{                                                                                       
Globular Cluster &\quad Pulsar &  RA (J2000)    &  Dec (J2000)      & $\delta$~RA$$     & $\delta$~dec$$     & Net count rate  & $P_{\rm coincide}$ \\
   {}            &  {}         &    h m s       &     d m s         &       arcsec      &         arcsec     & $10^{-4}$~cts/s &      $\%$           }    
\startdata                                                                                                                                                        
47\,Tucanae    & J0024-7204C  & 00 23 50.364  &   -72 04 31.54     &      0.308        &          0.226     &  $1.4\pm0.3$     &        0.018       \\
 {}            & J0024-7204D  & 00 24 13.882  &   -72 04 43.84     &      0.242        &          0.213     &  $3.4\pm0.4$     &        0.013       \\
 {}            & J0024-7204E  & 00 24 11.107  &   -72 05 20.27     &      0.249        &          0.214     &  $4.4\pm0.4$     &        0.014       \\
 {}            & J0024-7204F  & 00 24 03.936  &   -72 04 42.50     &      0.231        &          0.212     &  $6.1\pm0.6$     &        0.013       \\
 {}            & J0024-7204G  & 00 24 07.943  &   -72 04 39.65     &      0.210        &          0.210     &  $5.5\pm0.5$     &        0.011       \\
 {}            & J0024-7204H  & 00 24 06.712  &   -72 04 06.95     &      0.256        &          0.214     &  $2.7\pm0.3$     &        0.014       \\  
 {}            & J0024-7204I  & 00 24 07.936  &   -72 04 39.65     &      0.210        &          0.210     &  $5.5\pm0.5$     &        0.011       \\
 {}            & J0024-7204J  & 00 23 59.402  &   -72 03 58.93     &      0.262        &          0.219     &  $1.2\pm0.3$     &        0.015       \\    
 {}            & J0024-7204L  & 00 24 03.754  &   -72 04 56.87     &      0.210        &          0.210     &  $26.8\pm0.6$    &        0.011       \\  
 {}            & J0024-7204M  & 00 23 54.507  &   -72 05 30.77     &      0.286        &          0.219     &  $2.2\pm0.3$     &        0.016       \\    
 {}            & J0024-7204N  & 00 24 09.224  &   -72 04 28.95     &      0.252        &          0.214     &  $1.9\pm0.3$     &        0.013       \\
 {}            & J0024-7204O  & 00 24 04.614  &   -72 04 53.83     &      0.210        &          0.210     &  $9.4\pm0.6$     &        0.011       \\ 
 {}            & J0024-7204Q  & 00 24 16.518  &   -72 04 25.26     &      0.256        &          0.214     &  $2.5\pm0.3$     &        0.014       \\    
 {}            & J0024-7204R  & 00 24 07.551  &   -72 04 50.37     &      0.210        &          0.210     &  $5.4\pm0.4$     &        0.011       \\     
 {}            & J0024-7204S  & 00 24 03.936  &   -72 04 42.50     &      0.231        &          0.212     &  $6.1\pm0.5$     &        0.013       \\        
 {}            & J0024-7204T  & 00 24 08.552  &   -72 04 38.99     &      0.285        &          0.217     &  $0.8\pm0.2$     &        0.016       \\     
 {}            & J0024-7204U  & 00 24 09.873  &   -72 03 59.71     &      0.246        &          0.213     &  $3.1\pm0.4$     &        0.014       \\     
 {}            & J0024-7204W  & 00 24 06.035  &   -72 04 49.17     &      0.214$^{b}$  &        0.210$^{b}$ &  $30.9\pm1.2^d$   &       0.012       \\     
 {}            & J0024-7204Y  & 00 24 01.454  &   -72 04 41.83     &      0.251$^{b}$  &        0.215$^{b}$ &  $1.3\pm0.3$     &        0.014       \\\\[-1ex]
M28 (NGC\,6626)& J1824-2452A & 18 24 32.007   &   -24 52 10.49     &      0.210        & 0.219& $\!\!\!\!\!\!\!\!280.8 \pm 5.8$&        0.024       \\          
  {}           & J1824-2452G & 18 24 31.591   &   -24 52 17.49     &      0.226        &         0.341      &   $2.9 \pm 0.6$  &        0.036       \\          
  {}           & J1824-2452H & 18 24 31.591   &   -24 52 17.49     &      0.259        &         0.355      &   $2.0 \pm 0.7$  &        0.027       \\\\[-1ex]   
NGC\,6440       & J1748-2021B & 17 48 52.953   & -20 21 38.86      &       0.214       &         0.287      & variable$^{a}$   &        0.122       \\\\[-1ex]  
M62 (NGC\,6266) & J1701-3006B & 17 01 12.670   & -30 06 49.04      &       0.225       &         0.228      &   $9.0 \pm 1.3$  &        0.048       \\          
    {}          & J1701-3006C & 17 01 12.867   & -30 06 59.44      &       0.237       &         0.232      &   $2.7 \pm 0.9$  &        0.051       \\\\[-1ex]  
NGC\, 6752      & J1911-6000C & 19 11 05.556   & -60 00 59.68      &       0.297       &         0.248      &   $3.3 \pm 0.7$  &        0.007       \\          
                & J1910-5959D & 19 10 52.417   & -59 59 05.45      &       0.304       &         0.228      &   $7.4 \pm 1.1$  &        0.007       \\\\[-1ex]  
M4 (NGC\,6121)  & B1620-26    & 16 23 38.222   & -26 31 53.77      &       0.231       &         0.264      &   $4.3 \pm 1.1$  &        0.004       \\\\[-1ex]  
M71 (NGC\,6838) & J1953+1846A & 19 53 46.424   & 18 47 04.91       & 0.231$^{b}\!\!\!\!$&      0.238$^{b}$  &   $5.6 \pm 1.0$  &        0.016       \\\\[-1ex] 
Terzan 5        & J1748-2446  & 17 48 05.048   & -24 46 41.10      & 0.220$^{b}\!\!\!\!$&      0.219$^{b}$  &  $30.5 \pm 2.9^{c}$ &     0.008       \\\\[-1ex] 
NGC\,6397       & J1740-5340  & 17 40 44.589   & -53 40 40.90      &        0.218       &         0.235     &  variable$^{a}$  &        0.064       \\\\[-1ex]\hline         
\enddata                                                                                                                                                            

\tablecomments{
%
%
{\bf a.)} The net counting rates for PSR~J1748-2021B from observations in 2000-07-04 and 2003-06-27 are                  
$(16.8\pm2.8)\times 10^{-4}$ cts/s and $(50.3\pm4.6)\times 10^{-4}$ cts/s, respectively.  The net counting 
rates for PSR~J1740-5340 from observations in 2000-07-31, 2002-05-13, 2002-06-30, 2007-06-22, 2007-07-16 are 
$(13.1\pm1.7)\times 10^{-4}$ cts/s, $(17.4\pm 2.5)\times 10^{-4}$ cts/s, $(28.9\pm3.3)\times 10^{-4}$ cts/s,
$(22.6\pm1.6)\times 10^{-4}$ cts/s and $(29.2\pm1.4)\times 10^{-4}$ cts/s, respectively.                                                                                                                  
{\bf b.)} Positional uncertainty of the X-ray counterpart only. Errors for the radio pulsar timing position 
are unpublished.    
{\bf c.)} Based on observation ID.~3798 only.
{\bf d.)} Source exhibits variability at the binary period, count rate averaged over one orbit.
}                                                                                 
\end{deluxetable}

\clearpage

\begin{deluxetable}{l l c c c c c c}
\tabletypesize{\scriptsize}
\tablewidth{0pc}
\tablecaption{Counting rate upper limits for X-ray emission from globular cluster millisecond pulsars. \label{t:upperlimits}\label{table6}}
\tablehead{
Globular Cluster &\quad Pulsar &  RA (J2000)    &  Dec (J2000)   & Counts$^{a}$ &  Exp.~Time &    3-$\sigma$ UL$^{b}$       &  $f_x^d/10^{-15}$ \\
     {}          &  {}         &    h m s       &     d m s      &     {}       &     sec    &     $10^{-4}$~cts/s          &  erg/s/cm$^2$ }
\startdata

   NGC\,6440     & J1748-2021A &  17 48 52.689  &   -20 21 39.7  &   16.3       &   47325.1  &           3.9                &     6.77   \\                 
      {}         & J1748-2021C &  17 48 51.173  &   -20 21 53.81 &    2.2       &     {}     &           2.5                &     4.34   \\               
      {}         & J1748-2021D &  17 48 51.647  &   -20 21 07.41 &    2.2       &     {}     &           2.5                &     4.34   \\       
      {}         & J1748-2021E &  17 48 52.800  &   -20 21 29.38 &    1.1       &     {}     &           2.1                &     3.65   \\
      {}         & J1748-2021F &  17 48 52.334  &   -20 21 39.33 &    1.1       &     {}     &           2.3                &     3.99   \\\\[-1ex]
   NGC\,6441     & J1750-3703A &  17 50 13.802  &   -37 03 10.95 &    2.2       &   2290.1   &          50.7                &     175.6      \\ 
                 & J1750-3703B &  17 50 12.177  &   -37 03 22.93 &    1.1       &     {}     &          47.0                &     162.8      \\
                 & J1750-3703C &  17 50 13.454  &   -37 03 05.58 &    5.4       &     {}     &          59.7                &     206.8      \\  
                 & J1750-3703D &  17 50 13.097  &   -37 03 06.37 &   15.2       &     {}     &          78.8                &     272.9      \\\\[-1ex]
   NGC\,6544     & J1807-2459A &  18 07 20.36   &   -24 59 52.6  &    3.3       &  16277.6   &           7.6                &     11.02      \\\\[-1ex]
   NGC\,6752     & J1910-5959B &  19 10 52.056  &   -59 59 00.86 &   14.1       &  67435.9   &           2.6                &     1.94    \\ 
                 & J1910-5959E &  19 10 52.157  &   -59 59 02.09 &    8.7       &     {}     &           2.3                &     1.72    \\\\[-1ex]
 M13 (NGC\,6205) & J1641+3627A &  16 41 40.880  &    36 27 15.44 &    2.2       &  54693.8   &           2.1                &      1.53     \\\\[-1ex]
M15 (NGC\,7078)  & B2127+11A   &  21 29 58.247  &    12 10 01.26 & 1029.35$^{c}$&  30851.3   &           34.2               &     66.11    \\
    {}           & B2127+11B   &  21 29 58.632  &    12 10 00.31 &    60.9      &    {}      &           9.7                &     18.75    \\ 
    {}           & B2127+11C   &  21 30 01.204  &    12 10 38.21 &    1.1       &    {}      &           3.5                &     6.77    \\
    {}           & B2127+11D   &  21 29 58.274  &    12 09 59.74 &  150.0$^{c}$ &    {}      &           14.1               &     27.25      \\ 
    {}           & B2127+11E   &  21 29 58.187  &    12 10 08.63 &    81.5      &    {}      &           10.9               &     21.07      \\ 
    {}           & B2127+11F   &  21 29 57.178  &    12 10 02.91 &     6.5      &    {}      &           4.6                &     8.89      \\
    {}           & B2127+11G   &  21 29 57.948  &    12 09 57.26 &    65.2      &    {}      &           9.9                &    19.04      \\
    {}           & B2127+11H   &  21 29 58.184  &    12 09 59.43 &   155.4$^{c}$&    {}      &           14.3               &    27.64      \\\\[-1ex] 
M28 (NGC\,6626)  & J1824-2452B &  18 24 32.545  &   -24 52 04.29 &    19.6      & 196713.4   &           1.0                &     1.13    \\
    {}           & J1824-2452C &  18 24 33.089  &   -24 52 13.57 &    21.7      &    {}      &           1.0                &     1.13    \\
    {}           & J1824-2452D &  18 24 31.812  &   -24 49 25.03 &    9.8       &    {}      &           0.8                &     0.90    \\ 
    {}           & J1824-2452E &  18 24 32.900  &   -24 52 12.00 &   14.1       &    {}      &           0.9                &     1.01    \\ 
    {}           & J1824-2452F &  18 24 32.733  &   -24 52 10.18 &    7.8       &    {}      &           0.7                &     0.79    \\
    {}           & J1824-2452I &  18 24 32.192  &   -24 52 14.66 &    31.5      &    {}      &           1.2                &     1.35    \\ 
    {}           & J1824-2452J &  18 24 32.422  &   -24 52 25.90 &   117.4$^{c}$&    {}      &           2.0                &     2.25    \\\\[-1ex]
M30 (NGC\,7099)  & J2140-2310A &  21 40 22.406  &   -23 10 48.79 &     6.5      &  49435.3   &           2.9                &     2.17     \\\\[-1ex]
 M3 (NGC\,5272)  & J1342+2822B &  13 42 11.087  &    28 22 40.14 &     3.3      &  29525.9   &           4.2                &    3.02     \\
    {}           & J1342+2822D &  13 42 10.200  &    28 22 36.00 &     1.1      &     {}     &           3.3                &    2.37     \\\\[-1ex]
M5 (NGC\,5904)   & B1516+02A   &  15 18 33.318  &    02 05 27.55 &     2.2      & 44656.1    &           2.6                &    1.94     \\
    {}           & B1516+02B   &  15 18 31.458  &    02 05 15.47 &     2.2      &    {}      &           2.6                &    1.94     \\\\[-1ex]
M62 (NGC\,6266)  & J1701-3006A &  17 01 12.513  &   -30 06 30.13 &     4.3      &  62266.2   &           2.1                &    2.57     \\\\[-1ex]
   Terzan 5     & J1748-2446A  &  17 48 02.255  &   -24 46 36.90 &     2.2      &  39343.8   &           3.0                &    7.88    \\
     {}         & J1748-2446C  &  17 48 04.540  &   -24 46 36.00 &     3.3      &      {}    &           3.1                &    8.14    \\\hline

\enddata

\tablecomments{\tiny 
$^a$The total counts were extracted from a circle of 1 arcsecond radius (corresponding to 92\% of the encircled energy) and 
then rescaled to the 100\%. All counting rates are based on ACIS-S3 but for NGC 6441 and NGC 7078 (M15) which are based on 
HRC-I observations.
$^b$The 3-$\sigma$ upper limits were computered by using $C=0.5 \times (S/N)^2+(S/N) \times \sqrt{cts+0.25 \times (S/N)^{2}}$ ,
in which $S/N=3$ is the signal-to-noise ratio and $cts$ the counts obtained at the pulsar position. 
$^c$Upper limit dominated by a bright nearby source (cf.~Figure \ref{figure3}).
$^d$Unabsorbed Flux in the $0.3-8.0$ keV band.
%
%
}
\end{deluxetable}

\clearpage

\begin{deluxetable}{clc}
\tabletypesize{\scriptsize}
\tablewidth{0pc}
\tablecaption{Ephemeris used in the timing analysis of PSR J1824-2452A, J0023-7204D, J0023-7204O and J0023-7204R. \label{t:ephemeris}\label{table7}}
\tablehead{ \colhead{Pulsar} & {Parameter} & {Value} \\
}
\startdata
J1824-2452A &  MJD Range                                &  50351-52610          \\
   {}       &  $t_0$ (MJD)                              &  51468.0              \\
   {}       &  R.A. (J2000)                             &  18h24m32.s008345     \\
   {}       &  DEC.  (J2000)                            &  $-$24d52m10.s758586  \\
   {}       &  $\phi$($t_0$)                            &  0.6316               \\
   {}       &  $\nu$ (\perval{s}{-1})                   &  327.40564101150(1)   \\
   {}       &  $\dot{\nu}\,$(\ee{-12} \perval{s}{-2})   &  $-0.1735080(1)$      \\
   {}       &  $\ddot{\nu}\,$ (\ee{-24}\perval{s}{-3})  &  0.66(1)              \\
   {}       &  Time System                              &  DE405                \\\hline\\[-1ex]
J0024-7204D &  MJD Range                                &  48464-52357          \\
   {}       &  $t_0$ (MJD)                              &  51600                \\
   {}       &  R.A. (J2000)                             &  00h24m13.s87934(7)   \\
   {}       &  DEC.  (J2000)                            &  $-$72d04m43.s8405(3) \\
   {}       &  $\nu$ (\perval{s}{-1})                   &  186.651669856838(6)  \\
   {}       &  $\dot{\nu}\,$(\ee{-15} \perval{s}{-2})   &  0.1195(2)            \\
   {}       &  Time System                              &      DE200            \\\hline\\[-1ex]
J0024-7204O &  MJD Range                                &  50683-52357          \\
   {}       &  $t_0$ (MJD)                              &  51600                \\
   {}       &  R.A. (J2000)                             &  00h24m04.s6512(1)    \\
   {}       &  DEC.  (J2000)                            &  $-$72d04m53.s7552(5) \\
   {}       &  $\nu$ (\perval{s}{-1})                   &  378.30878836037(3)   \\
   {}       &  Time System                              &      DE200            \\
   {}       &  $P_b$(d)                                 &  0.1359743050(4)      \\
   {}       &  $a_p/c\, \sin(i)$(s)                     &  0.045151(2)          \\
   {}       &  $T_{asc}$(MJD)                           &  51600.0757554(6)     \\
   {}       &  $e$                                      &  $< 0.00016$          \\\hline\\[-1ex]
J0024-7204R &  MJD Range                                &                       \\
   {}       &  $t_0$ (MJD)                              &  51000                \\
   {}       &  R.A. (J2000)                             &  00h24m07.s(6)        \\
   {}       &  DEC.  (J2000)                            &  $-$72d04m50.s(1)     \\
   {}       &  $\nu$ (\perval{s}{-1})                   &  287.3181(0)          \\
   {}       &  Time System                              &       DE200           \\
   {}       &  $P_b$(d)                                 &  0.066(2)             \\
   {}       &  $a_p/c\, \sin(i)$(s)                     &  0.33(4)              \\
   {}       &  $T_{asc}$(MJD)                           &  50742.636(5)         \\
   {}       &  $e$                                      &  0.0                  \\
\enddata
\tablecomments{Ephemeris for J1824-2452A from Rots (2006) and references
therein. $t_0$ corresponds to phase zeor. $\phi$($t_0$) is the phase
of the main radio puls at $t_0$. Ephemeris of J0024-7204\,D,O are from Freire et al.~(2003) and 
for J0024-7204R from Camilo et al.~(2000).}
\end{deluxetable}

\clearpage

\begin{deluxetable}{l l c c c c c c}
\tabletypesize{\scriptsize}
\tablewidth{0pc}
\tablecaption{Spectral parameters and luminosities of X-ray detected globular cluster millisecond pulsar counterparts.
\label{t:gc_psr_spec}\label{table8}}
\tablehead{
 Cluster           &     Pulsar               & $N_{H}/10^{21}$   &  Model$^{a}$  & $\Gamma/kT (KeV)$      & $\chi^{2}_{\nu}$ (d.o.f) & log~$L_\mathrm{0.3-8~keV}$\,$^{b}$ & log~$L_\mathrm{0.1-2.4~keV}$\,$^{b}$\\
   {}              &       {}                 &  cm$^{-2}$        &       {}      &     {}                 &              &          ergs~s$^{-1}$                 &             ergs~s$^{-1}$         }
\startdata
   47\,Tucanae    & J0024-7204C               &      0.1          &       BB      & $0.24^{+0.05}_{-0.05}$ &  0.63(5)     &  $30.20^{+0.09}_{-0.12}$  & $30.22^{+0.09}_{-0.10}$  \\\\[-1ex]
    {}            & J0024-7204D               &      0.1          &       BB      & $0.21^{+0.04}_{-0.03}$ &  0.89(11)    &  $30.40^{+0.13}_{-0.14}$  & $30.44^{+0.11}_{-0.13}$ \\\\[-1ex]         
    {}            & J0024-7204E               &      0.1          &       BB      & $0.16^{+0.02}_{-0.02}$ &  0.67(8)     &  $30.56^{+0.10}_{-0.10}$  & $30.62^{+0.09}_{-0.07}$ \\\\[-1ex]       
    {}            & J0024-7204F/S             &      0.1          &       BB      & $0.21^{+0.01}_{-0.01}$ &  1.27(19)    &  $30.87^{+0.09}_{-0.08}$  & $30.91^{+0.10}_{-0.12}$ \\\\[-1ex]
    {}            & J0024-7204G/I             &      0.1          &       BB      & $0.24^{+0.04}_{-0.03}$ &  1.20(10)    &  $30.66^{+0.15}_{-0.12}$  & $30.67^{+0.12}_{-0.14}$ \\\\[-1ex]
    {}            & J0024-7204H               &      0.1          &       BB      & $0.19^{+0.08}_{-0.05}$ &  1.19(7)     &  $30.31^{+0.21}_{-0.17}$  & $30.35^{+0.18}_{-0.11}$ \\\\[-1ex]
    {}            & J0024-7204J               &      0.1          &     BB+PL     & $0.16^{+0.06}_{-0.04}~^{c}$/$0.61^{+0.55}_{-0.26}~^{d}$& 0.72(10) &  $31.06^{+0.27}_{-0.25}$  & $30.70^{+0.29}_{-0.26}$ \\\\[-1ex]
    {}            & J0024-7204L               &      0.1          &     BB+BB     & $0.15^{+0.02}_{-0.02}$/$1.95^{+0.58}_{-0.51}$  &  1.47(21) &  $31.60^{+0.28}_{-0.21}$  & $31.17^{+0.30}_{-0.26}$ \\\\[-1ex]
    {}            & J0024-7204M               &      0.1          &       BB      & $0.14^{+0.04}_{-0.04}$ &  0.99(4)     &  $30.49^{+0.14}_{-0.11}$  & $30.59^{+0.13}_{-0.13 }$ \\\\[-1ex]
    {}            & J0024-7204N               &      0.1          &       BB      & $0.48^{+0.18}_{-0.16}$ &  1.03(7)     &  $30.32^{+0.17}_{-0.21}$  & $30.20^{+0.28}_{-0.34}$ \\\\[-1ex]
    {}            & J0024-7204O               &      0.1          &     BB+PL     & $0.15^{+0.02}_{-0.02}~^{c}$/$1.33^{+0.24}_{-0.22}~^{d}$ &  1.06(10)  &  $31.34^{+0.31}_{-0.25}$  & $31.08^{+0.29}_{-0.18}$ \\\\[-1ex]
    {}            & J0024-7204Q               &      0.1          &       BB      & $0.30^{+0.11}_{-0.08}$ &  1.19(6)     &  $30.14^{+0.17}_{-0.13}$  & $30.14^{+0.19}_{-0.15}$ \\\\[-1ex]
    {}            & J0024-7204R               &      0.1          &     BB+PL     & $0.18^{+0.04}_{-0.03}~^{c}$/$1.51^{+0.67}_{-0.65}~^{d}$  &  1.71(16)  &  $31.44^{+0.27}_{-0.31}$  & $31.28^{+                                                                                    0.27}_{-0.29}$ \\\\[-1ex]
    {}            & J0024-7204T               &      0.1          &       BB      & $0.26^{+0.09}_{-0.07}$ & 0.99(5)      &  $30.18^{+0.13}_{-0.15}$  & $30.19^{+0.16}_{-0.18}$ \\\\[-1ex]
    {}            & J0024-7204U               &      0.1          &       BB      & $0.26^{+0.03}_{-0.06}$ & 1.00(10)     &  $30.36^{+0.08}_{-0.13}$  & $30.37^{+0.09}_{-0.11}$ \\\\[-1ex]
    {}            & J0024-7204W               &      0.1          &     BB+PL     & $0.06^{+0.01}_{-0.03}~^{c}$/$1.72^{+0.16}_{-0.17}~^{d}$  &  0.97(19) &  $31.39^{+0.12}_{-0.13}$  & $31.39^{+0.11}_{-0.13}$ \\\\[-1ex]
    {}            & J0024-7204Y               &      0.1          &       BB      & $0.14^{+0.09}_{-0.05}$ & 1.73(5)      &  $30.34^{+0.16}_{-0.15}$  & $30.43^{+0.15}_{-0.11}$ \\\\[-2ex]\hline\\[-2ex]

  M28 (NGC 6626)  & J1824-2452A               &     $2.2\pm0.2  $ &       PL      & $1.13^{+0.03}_{-0.04}$ &  1.00(168)   & 33.13$^{+0.05}_{-0.03}$   & 32.58$^{+0.03}_{-0.03}$  \\\\[-1ex]
                  & J1824-2452G               &      2.2          &       PL      &$2.7^{+0.4}_{-0.5}$     &  0.83(12)    & 31.14$^{+0.07}_{-0.08}$   & 31.49$^{+0.29}_{-0.34}$  \\\\[-2.5ex]
                  &                           &      2.2          &       BB      &$0.3^{+0.1}_{-0.1}$     &  1.01(12)    & 30.52$^{+0.15}_{-0.24}$   & 30.48$^{+0.28}_{-0.36}$  \\\\[-1ex]
                  & J1824-2452H               &      2.2          &       PL      & $0.7^{+0.3}_{-0.4}$    &  0.77(13)    & 31.20$^{+0.36}_{-0.30}$   & 30.53$^{+0.18}_{-0.16}$  \\\\[-2.5ex]
                  &                           &      2.2          &       BB      & $1.1^{+0.3}_{-0.2}$    &  0.73(13)    & 31.12$^{+0.22}_{-0.18}$   & 30.49$^{+0.35}_{-0.20}$  \\\\[-2ex]\hline\\[-2ex]
 NGC 6440         & J1748-2021B (\#947)       &      5.9          &       PL      & $1.6^{+0.7}_{-0.5}$    &  1.49(7)     & 32.32$^{+0.46}_{-0.39}$   & 32.10$^{+0.44}_{-0.28}$  \\\\[-2.5ex]
                  & J1748-2021B (\#3799)      &      5.9          &       PL      & $1.4^{+0.2}_{-0.2}$    &  0.93(14)    & 32.89$^{+0.16}_{-0.18}$   & 32.54$^{+0.10}_{-0.12}$  \\\\[-2ex]\hline\\[-2ex]
 M62 (NGC 6266)   & J1701-3006B               &      2.6          &       PL      & $2.1^{+0.3}_{-0.3}$    &  0.93(13)    & 31.95$^{+0.14}_{-0.13}$   & 31.98$^{+0.21}_{-0.18}$  \\\\[-2.5ex]
                  &    {}                     &      2.6          &       BB      & $0.5^{+0.1}_{-0.1}$    &  1.22(13)    & 31.69$^{+0.38}_{-0.26}$   & 30.52$^{+0.43}_{-0.27}$  \\\\[-1ex]
                  & J1701-3006C               &      2.6          &       PL      & $1.7^{+0.9}_{-0.9}$    &  0.10(3)     & 31.49$^{+0.55}_{-0.37}$   & 31.33$^{+0.50}_{-0.32}$  \\\\[-2ex]\hline\\[-2ex]
 NGC 6752         & J1911-6000C               &      0.2          &       PL      & $1.9^{+0.8}_{-0.8}$    &  0.98(4)     & 30.70$^{+0.39}_{-0.23}$   & 30.62$^{+0.45}_{-0.26}$  \\\\[-2.5ex]
                  &    {}                     &      0.2          &       BB      & $0.3^{+0.1}_{-0.1}$    &  0.56(4)     & 30.40$^{+0.36}_{-0.32}$   &  30.62$^{+0.39}_{-0.30}$ \\\\[-1ex]

                  & J1910-5959D               &      0.2          &       PL      & $2.6^{+0.5}_{-0.4}$    &  0.70(8)     & 30.99$^{+0.11}_{-0.06}$   & 31.26$^{+0.35}_{-0.27}$  \\\\[-2ex]\hline\\[-2ex]
 M4 (NGC 6121)    &  B1620-26                 &      2.0          &       PL      & $2.8^{+0.6}_{-0.5}$    &  1.05(8)     & 30.58$^{+0.09}_{-0.23}$   & 31.01$^{+0.31}_{-0.54}$  \\\\[-2.5ex]
                  &    {}                     &      2.0          &       BB      & $0.4^{+0.1}_{-0.1}$    &  1.23(8)     & 30.09$^{+0.23}_{-0.28}$   & 30.05$^{+0.19}_{-0.34}$  \\\\[-2ex]\hline\\[-2ex]

 M71 (NGC 6838)   & J1953+1846A               &      1.4          &       PL      & $1.9^{+0.5}_{-0.4}$    &  0.50(5)     & 31.15$^{+0.36}_{-0.30}$   & 31.10$^{+0.31}_{-0.54}$  \\\\[-2ex]\hline\\[-2ex]
 Terzan 5         & J1748-2446                &      12.0         &       PL      & $1.0^{+0.3}_{-0.2}$    &  0.44(10)    & 33.01$^{+0.19}_{-0.19}$   & 32.51$^{+0.19}_{-0.13}$  \\\\[-2.5ex]
                  &    {}                     &      12.0         &       BB      & $1.0^{+0.2}_{-0.1}$    &  0.70(10)    & 32.83$^{+0.17}_{-0.22}$   & 32.27$^{+0.21}_{-0.25}$  \\\\[-2ex]\hline\\[-2ex]

 NGC 6397         & J1740-5340 (\#79)         &      1.0          &       PL      & $1.8^{+0.3}_{-0.3}$    &  0.75/6      & 31.06$^{+0.32}_{-0.27}$   & 30.95$^{+0.31}_{-0.41}$  \\\\[-2.5ex]
                  &     {}                    &      1.0          &       BB      & $0.6^{+0.1}_{-0.1}$    &  0.67/6      & 30.82$^{+0.35}_{-0.38}$   & 30.63$^{+0.29}_{-0.33}$  \\\\[-1ex]
                  & J1740-5340 (\#2668)       &      1.0          &       PL      & $1.5^{+0.3}_{-0.3}$    &  0.75/8      & 31.08$^{+0.30}_{-0.44}$   & 30.81$^{+0.32}_{-0.38}$  \\\\[-2.5ex]
                  &     {}                    &      1.0          &       BB      & $0.6^{+0.1}_{-0.1}$    &  0.75/8      & 30.77$^{+0.37}_{-0.28}$   & 30.59$^{+0.19}_{-0.27}$  \\\\[-1ex]
                  & J1740-5340 (\#2669)       &      1.0          &       PL      & $1.5^{+0.2}_{-0.2}$    &  1.14/8      & 31.20$^{+0.17}_{-0.25}$   & 30.92$^{+0.16}_{-0.22}$  \\\\[-2.5ex]
                  &     {}                    &      1.0          &       BB      & $0.7^{+0.1}_{-0.1}$    &  1.28/8      & 30.97$^{+0.29}_{-0.20}$   & 30.70$^{+0.22}_{-0.31}$  \\\\[-1ex]
                  & J1740-5340 (\#7460)       & $0.9^{+1.0}_{-0.4}$\,$^{e}$& PL   & $1.4^{+0.1}_{-0.1}$    &  0.97/11     & 30.95$^{+0.15}_{-0.14}$   & 30.99$^{+0.12}_{-0.09}$  \\\\[-2.5ex]
                  & J1740-5340 (\#7461)       & $\lesssim 1.0$\,$^{e}$&   PL      & $1.7^{+0.4}_{-0.2}$    &  1.06/5      & 30.95$^{+0.26}_{-0.20}$   & 30.99$^{+0.13}_{-0.15}$  \\\hline
\enddata

\tablecomments{\tiny
Note: The meaning of the columns are as follows: Cols.~1 \&2  list the cluster and pulsar name, Cols.~3 to 5 list the hydrogen
column density, the spectral model and the best-fitted photon index or blackbody temperature, depending on the fitted model. Col.6 
lists the reduced $\chi^{2}$ together with the degrees of freedom given in brackets. Cols.~7 \& 8 list the unabsorbed X-ray luminosity 
in the $0.3-8.0$ and $0.1-2.4$\,keV energy ranges, respectively. Quoted errors indicate the 68\% confidence level for one parameter of interest.\\
\noindent
$^a$ PL $=$ power-law; BB $=$ blackbody \\
$^b$ Unabsorbed X-ray luminosities are computed at the distance of the GC (cf.~Table \ref{table3} and Harris 1996, updated 2003). \\
$^c$ The value indicates the blackbody temperature in the unit of keV. \\
$^d$ The value indicates the photon index inferred from the power-law model. \\
$^e$ The best-fitted $N_{H}$ values are consistent with the values of $\sim1.03\times 10^{21} \mathrm{cm^{-2}}$ which is inferred from the optical reddening of NGC 6397.}
\end{deluxetable}

\end{document}